\documentclass[final,3p,times,twocolumn]{elsarticle}
\usepackage{subfig}
\usepackage{siunitx}
\DeclareSIUnit\mwcm{\milli\W\per\square\cm}
\let\DeclareUSUnit\DeclareSIUnit
\let\US\SI
\DeclareUSUnit\inch{inch}
\usepackage{amssymb}
\usepackage{lineno}
\journal{Nuclear Instruments and Methods in Physics Research A}

\begin{document}

\begin{frontmatter}

  %% Title, authors and addresses use the tnoteref command within
  %% \title for footnotes; use the tnotetext command for the
  %% associated footnote; use the fnref command within \author or
  %% \address for footnotes; use the fntext command for the associated
  %% footnote; use the corref command within \author for corresponding
  %% author footnotes; use the cortext command for the associated
  %% footnote; use the ead command for the email address, and the form
  %% \ead[url] for the home page:
  %%
  %% \title{Title\tnoteref{label1}} \tnotetext[label1]{}
  %% \author{Name\corref{cor1}\fnref{label2}} \ead{email address}
  %% \ead[url]{home page} \fntext[label2]{} \cortext[cor1]{}
  %% \address{Address\fnref{label3}} \fntext[label3]{}

  \title{Thermal and hydrodynamic studies for micro-channel cooling for large area silicon sensors in high energy physics experiments}

  \author[label1]{Nils Flaschel} \author[label1]{Dario Ariza} \author[label1]{Sergio D{\'{i}}ez} \author[label2]{Marta Gerboles} \author[label1]{Ingrid-Maria Gregor} \author[label2]{Xavier Jorda} \author[label2]{Roser Mas} \author[label2]{David Quirion} \author[label1]{Kerstin Tackmann} \author[label2]{Miguel Ullan}

  \address[label1]{Deutsches Elektronen-Synchrotron, Notkestra\ss{}e 85, 22607 Hamburg} \address[label2]{Centro Nacional de Microelectr\'{o}nica, , Barcelona}

%%%%%%%%%%%%%%%%%%%%%%%%%%%%%%%%%%%%%%%%%%%%%%%%%%%%%%%%%%%%%%%%
%%%%%%%%%%%%%%%%%%%%%%%%%%%%%%%%%%%%%%%%%%%%%%%%%%%%%%%%%%%%%%%%
%%%%%%%%%%%%%%%%%%%%%%%%%%%%%%%%%%%%%%%%%%%%%%%%%%%%%%%%%%%%%%%%

  \begin{abstract}
    Micro-channel cooling initially aiming at small-sized high-power integrated circuits is being transferred to the field of high energy physics. Today's prospects of micro-fabricating silicon opens a door to a more direct cooling of detector modules. The challenge in high energy physics is to save material in the detector construction and to cool large areas.\\
    In this paper, we are investigating micro-channel cooling as a candidate for a future cooling system for silicon detectors in a generic research and development approach. The work presented in this paper includes the production and the hydrodynamic and thermal testing of a micro-channel equipped prototype optimized to achieve a homogeneous flow distribution. Furthermore, the device was simulated using finite element methods.
  \end{abstract}

%%%%%%%%%%%%%%%%%%%%%%%%%%%%%%%%%%%%%%%%%%%%%%%%%%%%%%%%%%%%%%%%
%%%%%%%%%%%%%%%%%%%%%%%%%%%%%%%%%%%%%%%%%%%%%%%%%%%%%%%%%%%%%%%%
%%%%%%%%%%%%%%%%%%%%%%%%%%%%%%%%%%%%%%%%%%%%%%%%%%%%%%%%%%%%%%%%

  \begin{keyword}
    micro-channel \sep cooling \sep HEP \sep silicon sensors \sep ANSYS \sep CFX \sep OpenFoam \sep microfluidics \sep heat sink \sep DESY \sep IMB-CNM \sep heat

    %% MSC codes here, in the form: \MSC code \sep code or \MSC[2008]
    %% code \sep code (2000 is the default)

  \end{keyword}

\end{frontmatter}



\section{Introduction}
\label{introduction}
As electronic devices are becoming smaller and smaller their power density is increasing and cooling becomes more important than ever before. The small dimensions of today's integrated circuits demand for a down scaling of heat sinks. The first cooling of electronics through micro-channel flows was accomplished in 1981 by Tuckerman and Pease \cite{Tuckerman1981}. The endeavor to achieve a better understanding of heat and mass transfer through micro-channels has led to an ever growing number of approaches. Using varying channel geometries, different fluids and multi-phase flows helped to increase the efficiency of micro-channel cooling and led to tailor-made solutions in various fields of applications \cite{AsselSakanovaa}.\\
The attractiveness of this technology to high energy physics (HEP) is less its capability to handle high power densities, but its potential to provide a direct and homogeneous cooling, to save material and to satisfy tight restrictions on the choice of materials, regarding radiation hardness and thermal expansion.\\
In HEP experiments like the ones of the LHC, the silicon detectors, which are positioned very close to the interaction point, receive considerable radiation doses. The sensors in tracking detectors, equipped with readout electronics, have to be kept at low temperatures around \SI{0}{\degreeCelsius} and in some cases much less \cite{Search2008}. This is mainly to keep the leakage current - introduced by radiation damage - low and to avoid thermal runaway. The cooling system typically adds a significant amount of material to the detector, leading to multiple scattering of charged particle tracks and conversions of photons into electron-positron pairs when passing through the material.\\
Saving material is essential to maintaining a good momentum resolution especially for low momentum particles. To do so, the heat sinks for the sensors, the readout electronics and the rest of the detector have to be within a certain radiation length, to limit multiple scattering and conversions. Moreover, the difference of the coefficient of thermal expansion (CTE) between sensor and heat sink must be kept small to avoid mechanical stress occurring by undergoing temperature cycles during production, installation and operation.\\
The micro-channel system presented in this paper is built upon a channel array etched in silicon, closed with a Pyrex layer and operated with a single-phase hydrofluoroether coolant \cite{hfe}. Pyrex was chosen for the purpose of the studies presented here, because it provides the ability to visually inspect the channel array for impurities and empty channels. A design has been developed to maintain a homogeneous distribution of the coolant across the micro-channel layout. For thermal characterization, heat was distributed homogeneously across the device. A design cooling mainly the parts exposed to a heat flow was already demonstrated in the ALICE ITS detector development \cite{alice} and a material reduction via thinning of the micro-channel device by the NA62 group \cite{Romagnoli2015}.\\
Besides the ALICE and the NA62 experiment, also LHCb \cite{Nomerotski} developed a micro-channel cooling solution for their application. The approaches of these experiments are similar to the one used in the present work, but with different layouts and coolants. The LHCb micro-channel cooling system is based on an evaporative CO$_{2}$ coolant and a channel design adopted to that mode of operation. The NA62 experiment on the other hand utilizes also a single-phase perfluorocarbon coolant, but a different flow distributing method by using a double layout design with two inlets and two outlets, different channel sizes and manifold design. The ALICE experiment is based on a small area cooling design. This work is especially interested in the homogeneous distribution of the coolant across a large area to maintain small temperature gradients.

%%%%%%%%%%%%%%%%%%%%%%%%%%%%%%%%%%%%%%%%%%%%%%%%%%%%%%%%%%%%%%%%
%%%%%%%%%%%%%%%%%%%%%%%%%%%%%%%%%%%%%%%%%%%%%%%%%%%%%%%%%%%%%%%%
%%%%%%%%%%%%%%%%%%%%%%%%%%%%%%%%%%%%%%%%%%%%%%%%%%%%%%%%%%%%%%%%

\section{Layout and calculations}
\label{Layout}
\begin{figure}[tb]
  \includegraphics[width=0.47\textwidth]{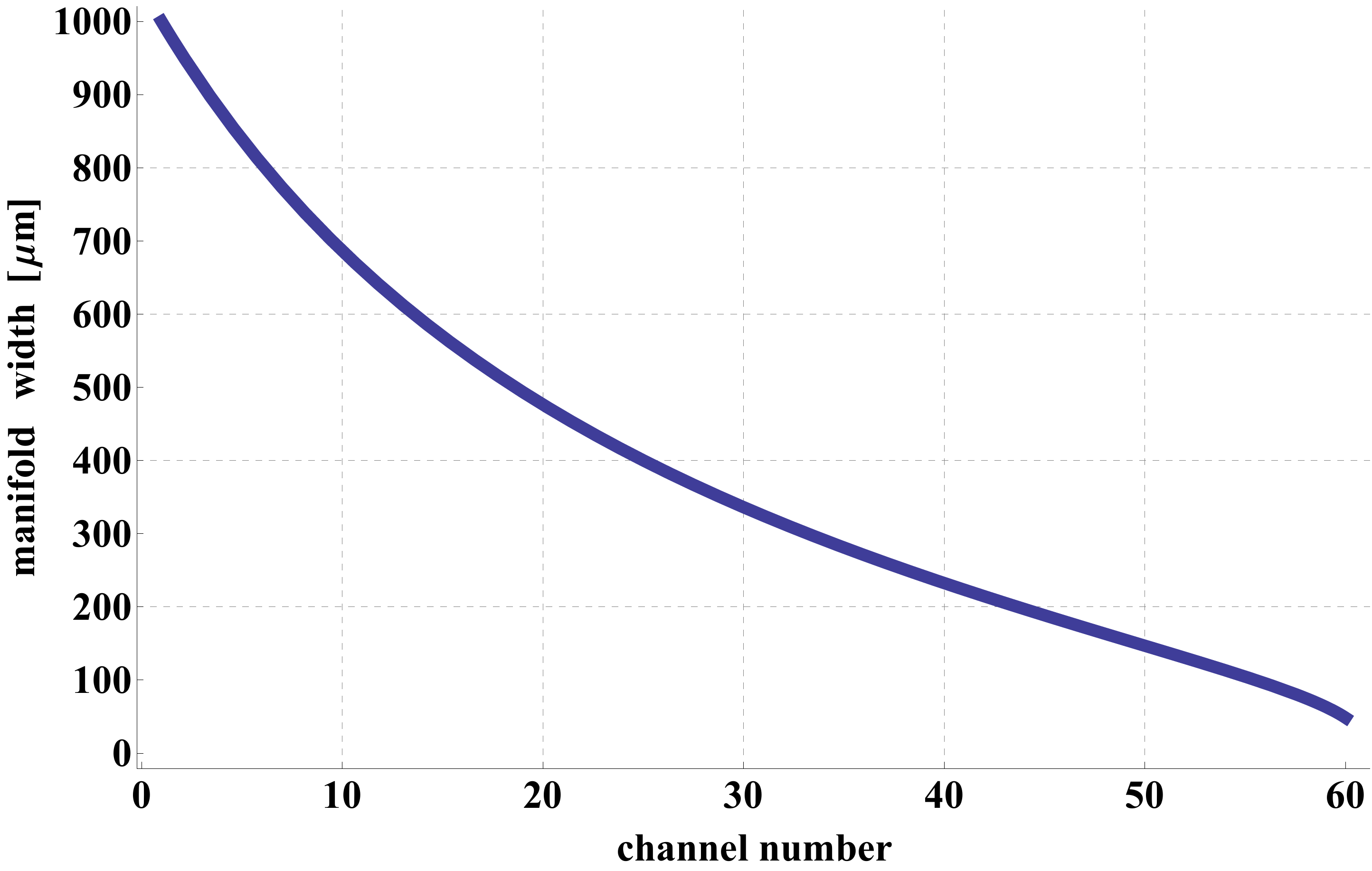}
  \caption{The width of the manifold according to equation \ref{eq:manifold}, for 60 channels and a starting width of \SI{1000}{\um} at the first entry to a smaller channel.}
  \label{fig:manifold}
\end{figure}

\begin{figure}[tb]
  \includegraphics[width=0.47\textwidth]{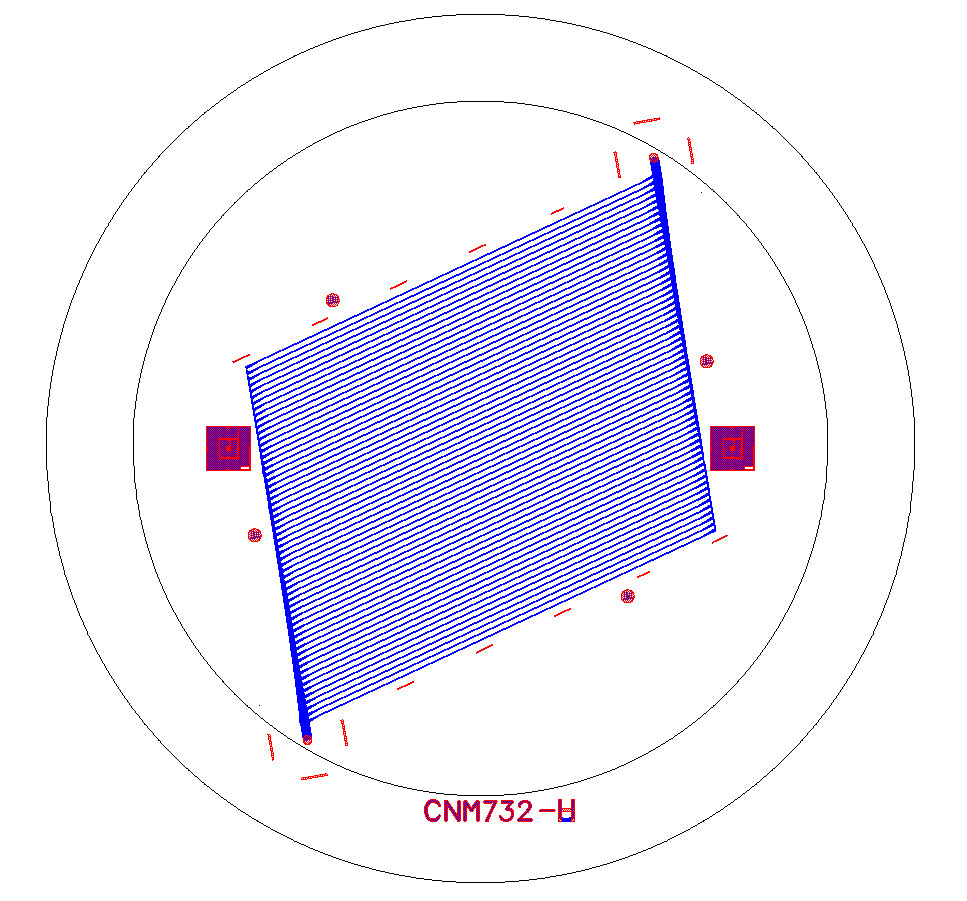}
  \caption{Layout of the micro-channels, projected on the outlines of a \US{4}{\inch} wafer.}
  \label{fig:layout}
\end{figure}

The layout of the micro-channel cooling device was designed to fit on a \US{4}{\inch} wafer. To keep the pressure drop as low as possible, a design was chosen based on manifolds connecting a number of smaller channels. One inlet and one outlet oppositely arranged are then connected to a cooling circuit. To provide a homogeneous flow through all $n$ channels, the pressure at each transition from manifold to smaller channel must be the same. From that assumption follows:
\begin{equation}
  p_{n}=p_{n+1},
  \label{pressureEquality}
\end{equation}
with $p_{n}$ the pressure in the manifold at the entrance to the $n$-th channel. The pressure must be adjusted by decreasing the width of the inlet manifold along the flow direction. The flow in the inlet manifold decreases with the number of channel entries along the way:
\begin{equation}
  Q_{n}=Q_{0}-n\frac{Q_{0}}{N},
\end{equation}
with $Q_{0}$ being the total flow through the device, $N$ the total number of channels and $Q_{n}$ the flow through the manifold at channel $n$. The pressure drop derived by the Navier-Stokes equation for a circular channel cross section is used as a good approximation:
\begin{equation}
  \Delta p=Q\frac{8\eta L}{\pi a^{4}},
\end{equation}
with $\eta$ the viscosity, $L$ the overall channel length and $a$ the characteristic dimension. The characteristic dimension of the manifold at channel $n$ is given by $a_{m}$
\begin{equation}
  a_{m}^{4}(n)=(1-n\frac{1}{N})a_{c}^{4}
\end{equation}
with $a_{c}$ the characteristic length of the manifold at the first channel entry. The characteristic length is chosen to be the hydraulic diameter which - in the case of a fully filled channel - is
\begin{equation}
  a = 2*\frac{w*h}{w+h},
\end{equation}
with $h$ the channel height and $w$ the width of the channel. The hydraulic diameter of the manifold at the entry to channel number $n$ is $a_{m}(n)$. This results in the width of the manifold at channel $n$ being
\begin{equation}
  w_{m}(n)=\frac{h\left(-\frac{w_{c}^{4}h^{4}(-1+n-N)}{(w_{c}+h)^{4}N}\right)^{\frac{1}{4}}}{-h+\left(-\frac{w_{c}^{4}h^{4}(-1+n-N)}{(w_{c}+h)^{4}N}\right)^{\frac{1}{4}}}
  \label{eq:manifold}
\end{equation}
with a total channel number of $N$. For technical reasons the manifold function was approximated by eleven linear sections.\\
For mechanical robustness of the \SI{300}{\um} to \SI{500}{\um} thick silicon wafer, the size of the channels was chosen to be \SI{100}{\um} $\times$ \SI{100}{\um} with a starting width of the manifold $w_{c}$ = \SI{1000}{\um} at the first channel entry. The structure has 60 parallel channels from inlet to outlet manifolds, with a pitch of \SI{675}{\um}. Figure \ref{fig:manifold} shows the shape of $w_{m}(n)$ for $n$ = 60 channels and a starting width of \SI{1000}{\um}.
The channels were designed with a \SI{15}{\degree} angle with respect to the perpendicular of the manifolds in order to avoid possible flow separations and limit resulting recirculating flows \cite{fluidFundamentals}. Inlet and outlet holes to the manifolds with \SI{1}{\mm} diameter were added. Finally the whole design was rotated by \SI{10}{\degree} with respect to the wafer axis to avoid coincidence of the channel etch direction with the silicon crystallographic planes, in order to avoid cracks or full wafer cleaving.\\
The channel ends were widened to prevent possible turbulence at sharp edges resulting in a structure shown in figure \ref{fig:layout}. Additional technical test structures, and alignment marks were added to facilitate processing.

%%%%%%%%%%%%%%%%%%%%%%%%%%%%%%%%%%%%%%%%%%%%%%%%%%%%%%%%%%%%%%%%
%%%%%%%%%%%%%%%%%%%%%%%%%%%%%%%%%%%%%%%%%%%%%%%%%%%%%%%%%%%%%%%%
%%%%%%%%%%%%%%%%%%%%%%%%%%%%%%%%%%%%%%%%%%%%%%%%%%%%%%%%%%%%%%%%
\section{Fabrication}
\label{Fabrication}
The fabrication of the micro-channel assembly structure was performed in the clean room of the Centro Nacional de Microelectronica (IMB-CNM, CSIC), Barcelona, Spain. The assembly was performed in two steps. In the first step, a silicon wafer was micro-machined by Deep Reactive Ion Etching (DRIE) in order to create the channels and the inlet and outlet in the silicon. In the second step, the micro-machined silicon wafer was bonded by an anodic process to a blank Pyrex wafer in order to seal the micro-channels and define the cooling structure inside the assembly.\\
\begin{figure}[tb]
  \includegraphics[width=0.47\textwidth]{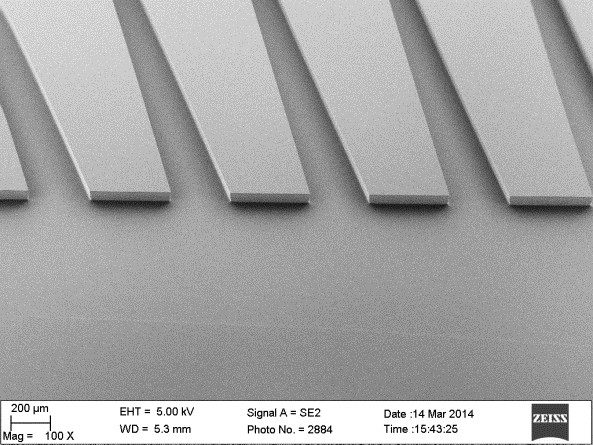}
  \caption{Electronic microscopy image of the channels produced by the DRIE process.}
  \label{fig:rem01_pic}
\end{figure}
\begin{figure}[tb]
  \includegraphics[width=0.47\textwidth]{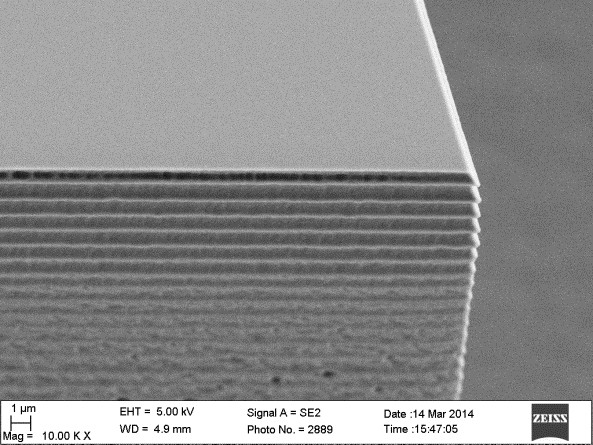}
  \caption{Detailed view of the scalloping effect at the micro-channel walls produced by DRIE process, note the smaller scale compared to figure \ref{fig:rem01_pic}.}
  \label{fig:rem02_pic}
\end{figure}
\begin{figure}[tb]
  \includegraphics[width=0.46\textwidth]{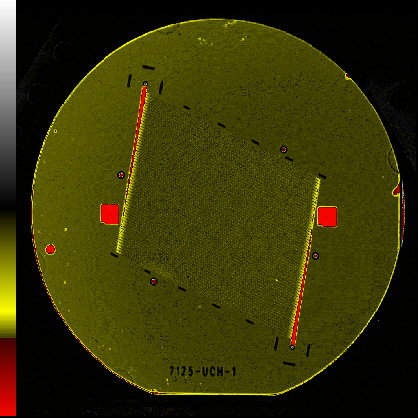}
  \caption{Scanning Acoustic Microscopy image of the assembly produced by anodic bonding.}
  \label{fig:acoustic_pic}
\end{figure}
\begin{figure}[tb]
  \includegraphics[width=0.46\textwidth]{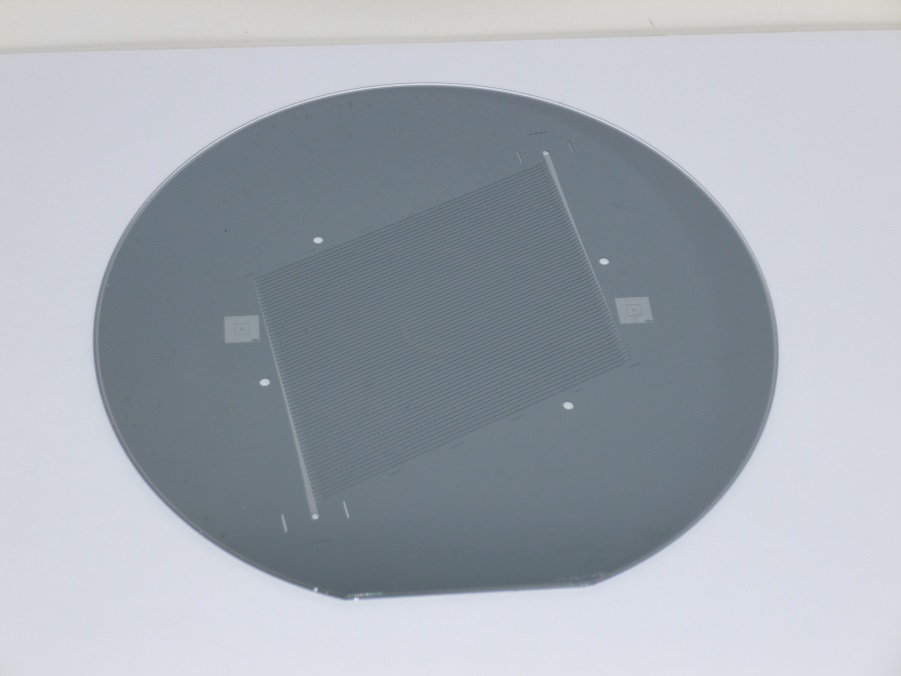}
  \caption{Optical image of the resulting assembly. The four white dots around the channel structure are alignment marks.}
  \label{fig:optical_order}
\end{figure}
\begin{figure}[tb]
  \includegraphics[width=0.4\textwidth]{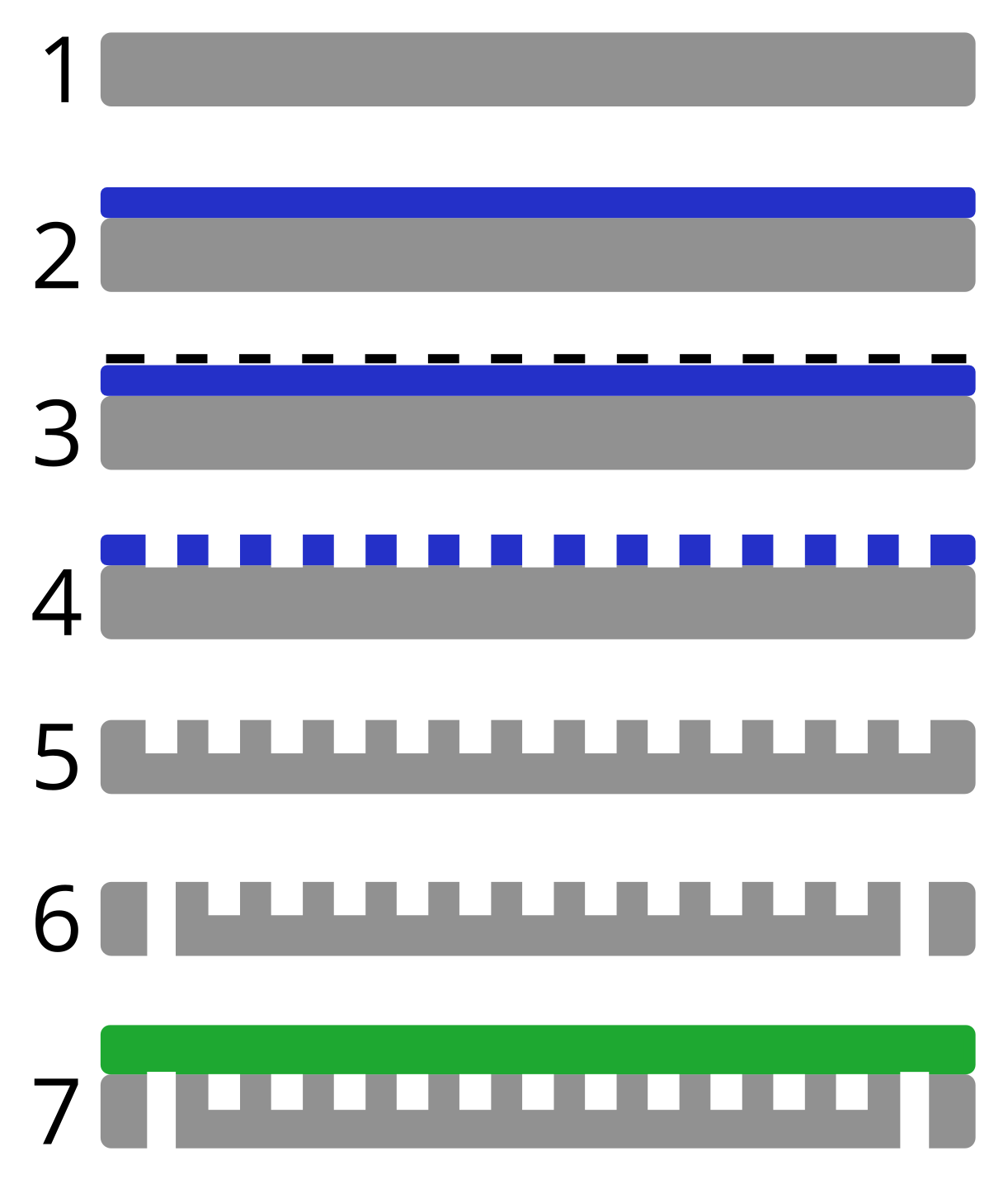}
  \caption{The basic steps of micro-channel fabrication. One side of blank wafer (1) is coated with a photo-lithographic resist (2). A mask of the layout is used (3) to transfer the channel design to the resist (4). The \SI{100}{\um} deep channels are etched into the silicon (5), the inlet and outlet holes are added (6) and closed with a Pyrex layer (7).}
  \label{fig:etching}
\end{figure}
As shown in figure \ref{fig:etching}, the process starts with a blank \US{4}{\inch} diameter, \SI{300}{\um} to \SI{500}{\um} thick, double-side polished silicon wafer. After a standard cleaning of the wafer, the layout previously defined was transferred to the wafer by a photo-lithography process. The photo-lithographic resist was used as mask during the following silicon etching process. Then, a \SI{100}{\um} etch was processed with an ALCATEL 601-E equipment. The etching was carried out by the Bosch process based on consecutive steps of silicon plasma etch and deposition of a passivation layer which protects the silicon walls, achieving in this way a very anisotropic etching and a high aspect ratio of the micro-channels \cite{laermer} \cite{plasma}. A Scanning Electron Microscopy (SEM) picture of the result of this etch and details of the resulting micro-channels can be seen in figure \ref{fig:rem01_pic} and \ref{fig:rem02_pic} where the characteristic rippled structure (“scalloping”) of the DRIE can be seen in the micro-channel walls.\\
After the micro-channels were created in one side of the wafer, a thermal oxide was grown at both wafer sides to protect the micro-channels from the next etching process. A metal mask was defined on the back side of the wafer by a second photolithographic step, leaving open the areas that have to be etched from the back side. This mask has to be aligned with the micro-channels at the other side of the wafer (double-sided alignment). The silicon oxide layer previously grown was then etched by a physical dry process in the open areas not covered with metal, and later the second DRIE etch was performed on the same open areas. In this process, the etching continues for the whole silicon thickness, creating through holes to define the inlet and outlet entries to the micro-channels. Finally, the metal mask was removed, the protecting oxide was wet etched, and a deep cleaning (RCA clean) was performed on the wafer to leave the bare silicon exposed for the next step.\\
In the second step of the assembly fabrication, a blank Pyrex wafer, \SI{500}{\um} thick, was used to close the micro-channels created in the silicon wafer by an anodic bonding process \cite{microfabrication} \cite{bonding}. The equipment used was a S\"USS MicroTec SB6e. For the anodic bonding a high potential difference (\SI{1000}{\V}) was applied between the two wafers which are in close contact. The high electric field across the wafers enhances the ion interchange needed to form the Si-SiO$_{2}$ bonding with a relatively low temperature ($<$ \SI{400}{\degreeCelsius}). The process creates a strong bond between the two wafers sealing the micro-channels.\\
The assembly was then analyzed by Scanning Acoustic Microscopy (SAM) with a Sonoscan GEN-5 equipment, which detects any voids between the wafers, to test the bond quality. An example of the SAM results can be seen in figure \ref{fig:acoustic_pic}. The micro-channels sealed in the interior of the assembly can be clearly identified together with the alignment marks besides them. Only a small negligible void, recognizable as red dot in figure \ref{fig:acoustic_pic}, can be seen at the very left edge far away from the micro-channels, probably due to some small particle on either of the two wafers before the process. In figure \ref{fig:optical_order} an optical image of the assembly can be seen where the micro-channels are visible through the transparent Pyrex wafer that seals them.

%%%%%%%%%%%%%%%%%%%%%%%%%%%%%%%%%%%%%%%%%%%%%%%%%%%%%%%%%%%%%%%%
%%%%%%%%%%%%%%%%%%%%%%%%%%%%%%%%%%%%%%%%%%%%%%%%%%%%%%%%%%%%%%%%
%%%%%%%%%%%%%%%%%%%%%%%%%%%%%%%%%%%%%%%%%%%%%%%%%%%%%%%%%%%%%%%%
\section{Simulation}
\label{simulation}
%%%%%%%%%%%%%%%%%%%%%%%%%%%%%%%%%%%%%%%%%%%%%%%%%%%%%%%%%%%%%%%%
%%%%%%%%%%%%%%%%%%%%%%%%%%%%%%%%%%%%%%%%%%%%%%%%%%%%%%%%%%%%%%%%
%%%%%%%%%%%%%%%%%%%%%%%%%%%%%%%%%%%%%%%%%%%%%%%%%%%%%%%%%%%%%%%%
\subsection{Simulation environment}
\label{simu_env}
Simulations are useful to study effects on the measurements due to changes in the setup such as using a different fluid, different boundary conditions, or can simply help to provide an understanding of a malfunctioning device.\\
The interFoam solver of the open source computational fluid dynamics software OpenFoam \cite{openfoam} was used to simulate the initial filling process of the micro-channels in a multi-phase approach. Furthermore the CFX package of the numeric simulation software ANSYS \cite{ansys} was used to build a 3d model to calculate flow and thermal properties of the micro-channel device in a single-phase approach. The layer structure of the micro-channel walls as visible in figure \ref{fig:rem02_pic} was not implemented in the simulation. The surrounding atmosphere was neglected as well, which means that there is no heat transfer through the wafer surface to the outside. Thermal properties were simulated modeling the channels in the silicon, closed with a Pyrex layer and ports attached to the inlet and the outlet.\\
The simulation of the above setup included flow calculations as well as thermal calculations, thus the mesh can be split into a fluid volume and a solid volume with different requirements. The volumes representing the fluid demand a much finer granularity than the solid parts which play only a role for the thermal calculations. An overall element size of \SI{10}{\um} to \SI{30}{\um} was chosen along the cross section of the smaller channels and an element size of \SI{800}{\um} along the channel length. The overall element size of the manifolds was kept between \SI{20}{\um} and \SI{70}{\um}. Three to four mesh layers were placed on all fluid boundary walls with a starting thickness of \SI{2}{\um} and a growth rate of 1.2. The solid compartments were meshed coarser, with element sizes ranging from \SI{800}{\um} for the wafer bulk material, down to \SI{100}{\um} at contact surfaces. This resulted in models with a number of elements ranging from a few million to over twenty million. The properties of the fluid in the simulation are kept close to the one in the experiment, with a density of around \SI{1500}{\kilogram\per\cubic\metre}, and a dynamic viscosity of \SI{0.58}{\gram\per\metre\per\second} at around \SI{20}{\degreeCelsius}. Because the maximum fluid velocity in the simulation was found to be around \SI{3}{\metre\per\second}, resulting in a maximum Reynolds number of around 1100, a laminar fluid model was chosen for the calculations.

%%%%%%%%%%%%%%%%%%%%%%%%%%%%%%%%%%%%%%%%%%%%%%%%%%%%%%%%%%%%%%%%
%%%%%%%%%%%%%%%%%%%%%%%%%%%%%%%%%%%%%%%%%%%%%%%%%%%%%%%%%%%%%%%%
%%%%%%%%%%%%%%%%%%%%%%%%%%%%%%%%%%%%%%%%%%%%%%%%%%%%%%%%%%%%%%%%
\subsection{Simulation results}
\label{simu_res}

\begin{figure}[tb!]
  \subfloat[]{\includegraphics[width=0.46\textwidth]{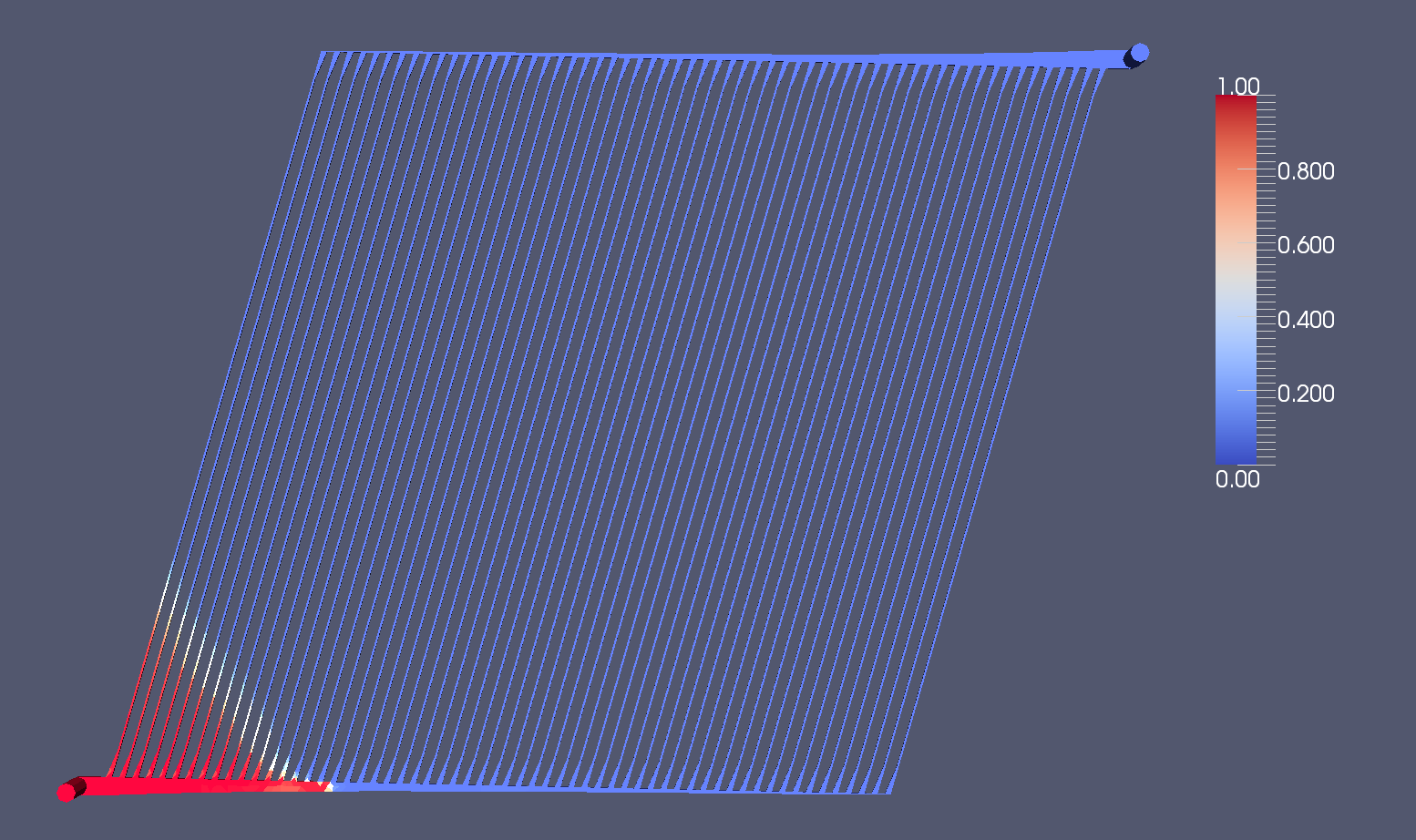}}\\
  \subfloat[]{\includegraphics[width=0.46\textwidth]{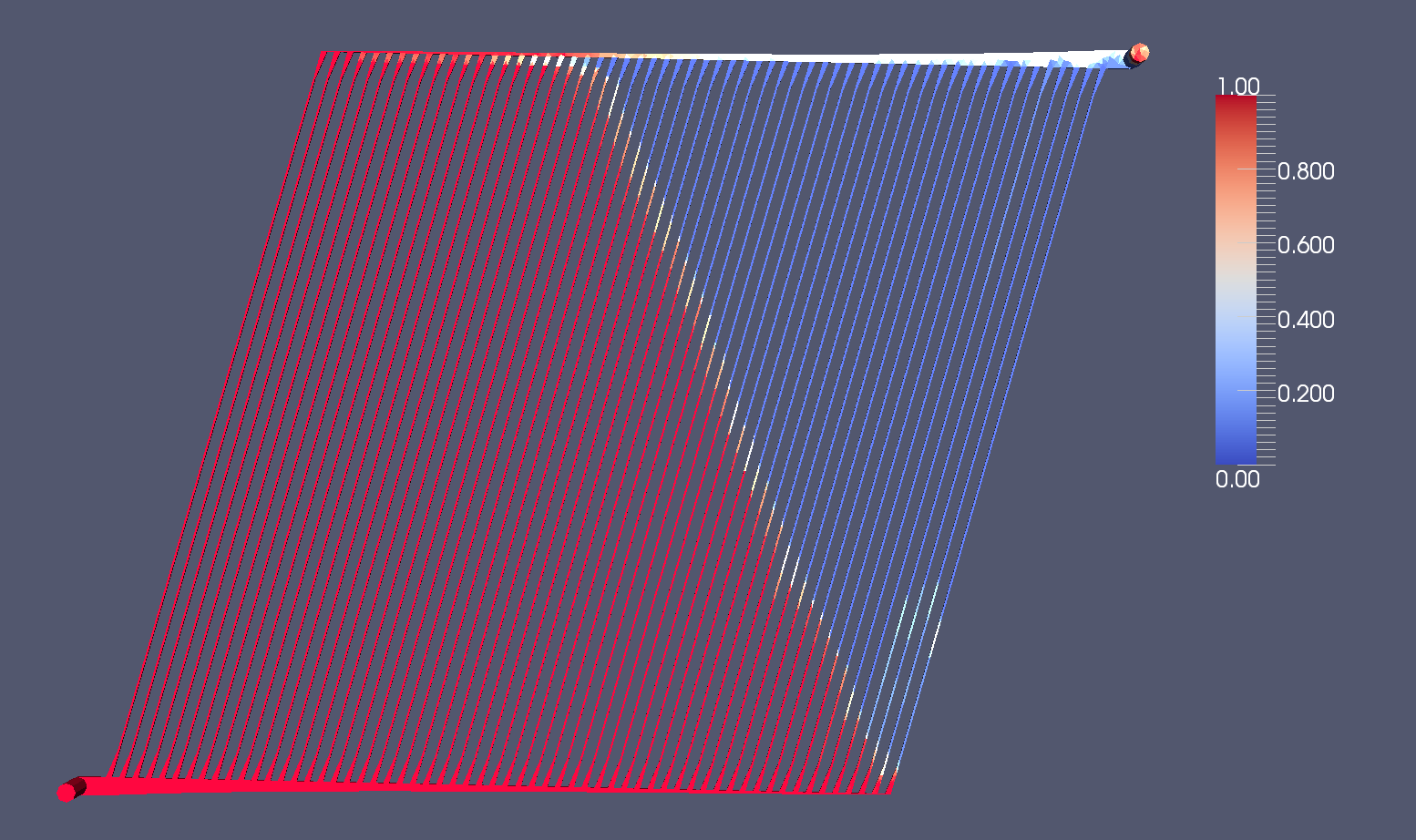}}\\
  \subfloat[]{\includegraphics[width=0.46\textwidth]{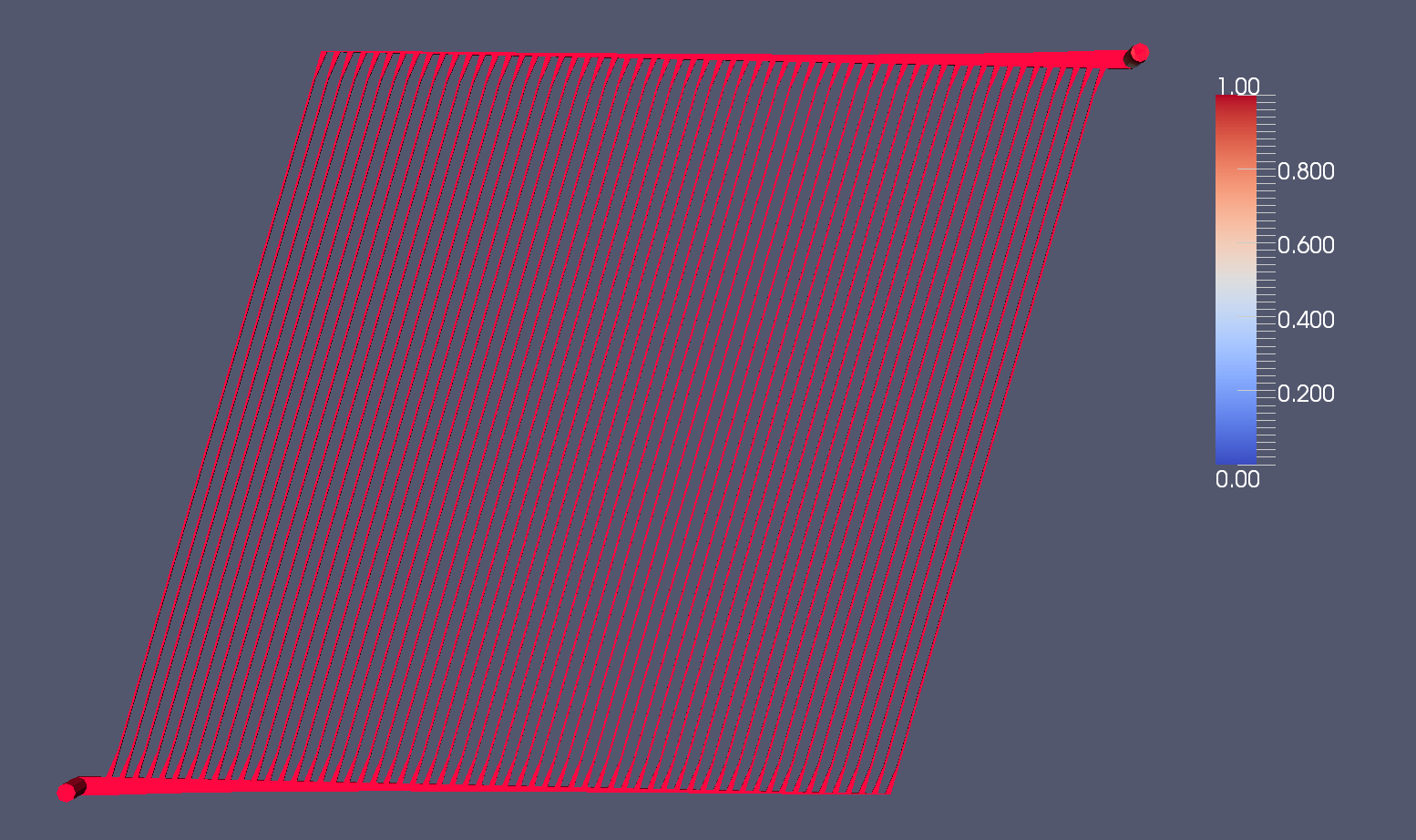}}
  \caption{The initial filling of the micro-channels simulated with OpenFoam. The red colored phase represents the fluid and the blue phase the air. The pressure was increased over time from \SIrange{0.1}{0.5}{\bar}, resulting in a flow rate of around \SIrange{1}{5}{\ml\per\minute}. Figure a$)$ shows the filling process after \SI{0.06}{\second}, figure b) after \SI{2.19}{\second} and c) after \SI{6}{\second}.}
  \label{fig:filing_03}
\end{figure}
Results of the simulated filling process are shown in figure \ref{fig:filing_03}. They show a good agreement with the behavior observed in the experiment by filming the filling process through the Pyrex. The homogeneous distribution of the fluid is very important to avoid dryouts due to insufficient flow in certain regions or empty channels due to air enclosures at channel entries, creating clogging \cite{Wyss2006a}, which can lead to additional pressure drops or complete malfunction.\\
\begin{table}[tb]
  \centering
  \caption{Calculated pressure values $p$ from the simulation of flow rates $Q$ between \SI{4}{\ml\per\minute} and \SI{40}{\ml\per\second}.}
  \begin{tabular}{|c|c||c|c|}
    \hline
    $Q$ [\si{\ml\per\minute}] & $p$ [\si{\bar}] & $Q$ [\si{\ml\per\minute}] & $p$ [\si{\bar}]\\
    \hline
    \hline
    4  & 0.5 & 24 & 4.3\\
    8  & 1.1 & 28 & 5.3\\
    12 & 1.8 & 32 & 6.4\\
    16 & 2.6 & 36 & 7.9\\
    20 & 3.4 & 40 & 9.1\\
    \hline
  \end{tabular}
  \label{tab:simuPressFlow}
\end{table}
Flow characteristics were calculated in the range of \SIrange{10}{40}{\ml\per\minute} with the flow properties of hydrofluoroether (HFE) \cite{hfe}. The results in table \ref{tab:simuPressFlow} show the pressure $p$ calculated at the inlet for a given flow rate $Q$.\\
After simulating the flow, the thermal properties of the solids were added to the simulation. The inlet temperature was set to \SI{19}{\degreeCelsius}. As mentioned previously, the convective and radiative heat transfer to the environment was not implemented in the simulation, thus the effective heat flux through the wafer differed from the experiment. The results of the thermal study are presented together with the experimental results in section \ref{exp_res}.

%%%%%%%%%%%%%%%%%%%%%%%%%%%%%%%%%%%%%%%%%%%%%%%%%%%%%%%%%%%%%%%%
%%%%%%%%%%%%%%%%%%%%%%%%%%%%%%%%%%%%%%%%%%%%%%%%%%%%%%%%%%%%%%%%
%%%%%%%%%%%%%%%%%%%%%%%%%%%%%%%%%%%%%%%%%%%%%%%%%%%%%%%%%%%%%%%%

\section{Measurements}
\label{experiment}
%%%%%%%%%%%%%%%%%%%%%%%%%%%%%%%%%%%%%%%%%%%%%%%%%%%%%%%%%%%%%%%%
%%%%%%%%%%%%%%%%%%%%%%%%%%%%%%%%%%%%%%%%%%%%%%%%%%%%%%%%%%%%%%%%
%%%%%%%%%%%%%%%%%%%%%%%%%%%%%%%%%%%%%%%%%%%%%%%%%%%%%%%%%%%%%%%%

\subsection{Experimental setup}
\label{exp_set}
\begin{figure}[tb]
  \includegraphics[width=0.46\textwidth]{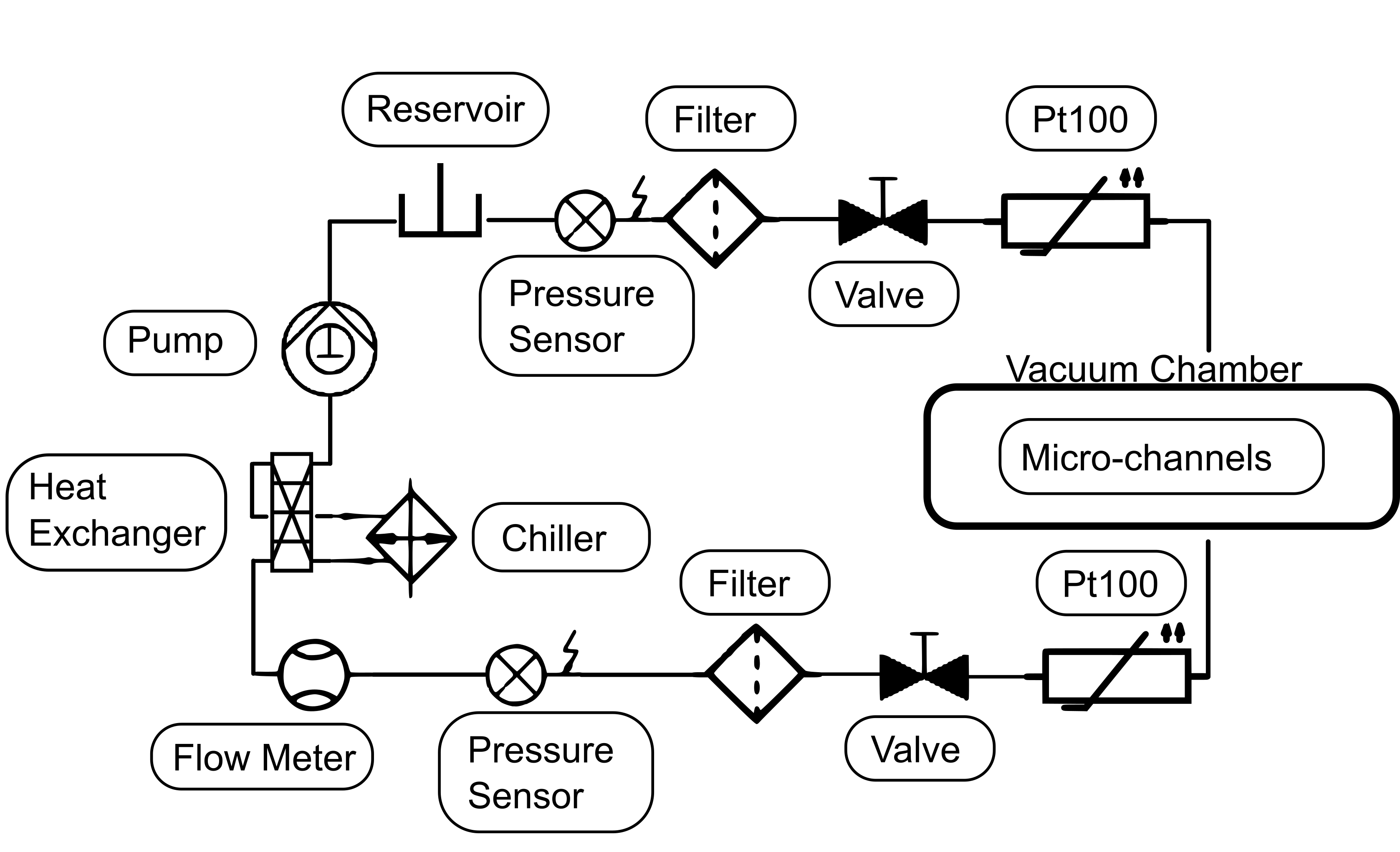}
  \caption{Sketch of the setup.}
  \label{fig:setup_pic}
\end{figure}
\begin{figure}[tb]
  \includegraphics[width=0.46\textwidth]{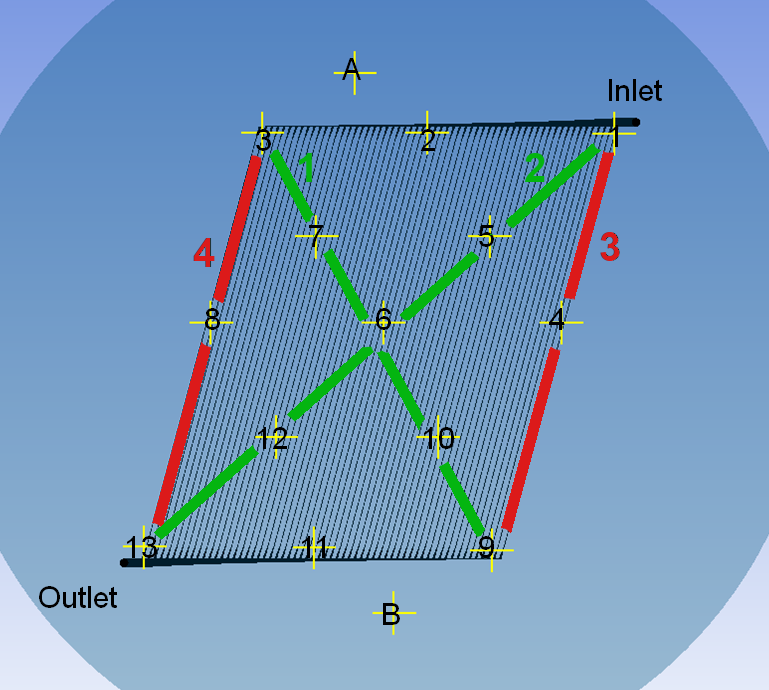}
  \caption{Pt100 temperature dependent resistors placed on the silicon surface.}
  \label{fig:sensor_order}
\end{figure}
\begin{figure}[tb]
  \includegraphics[width=0.46\textwidth]{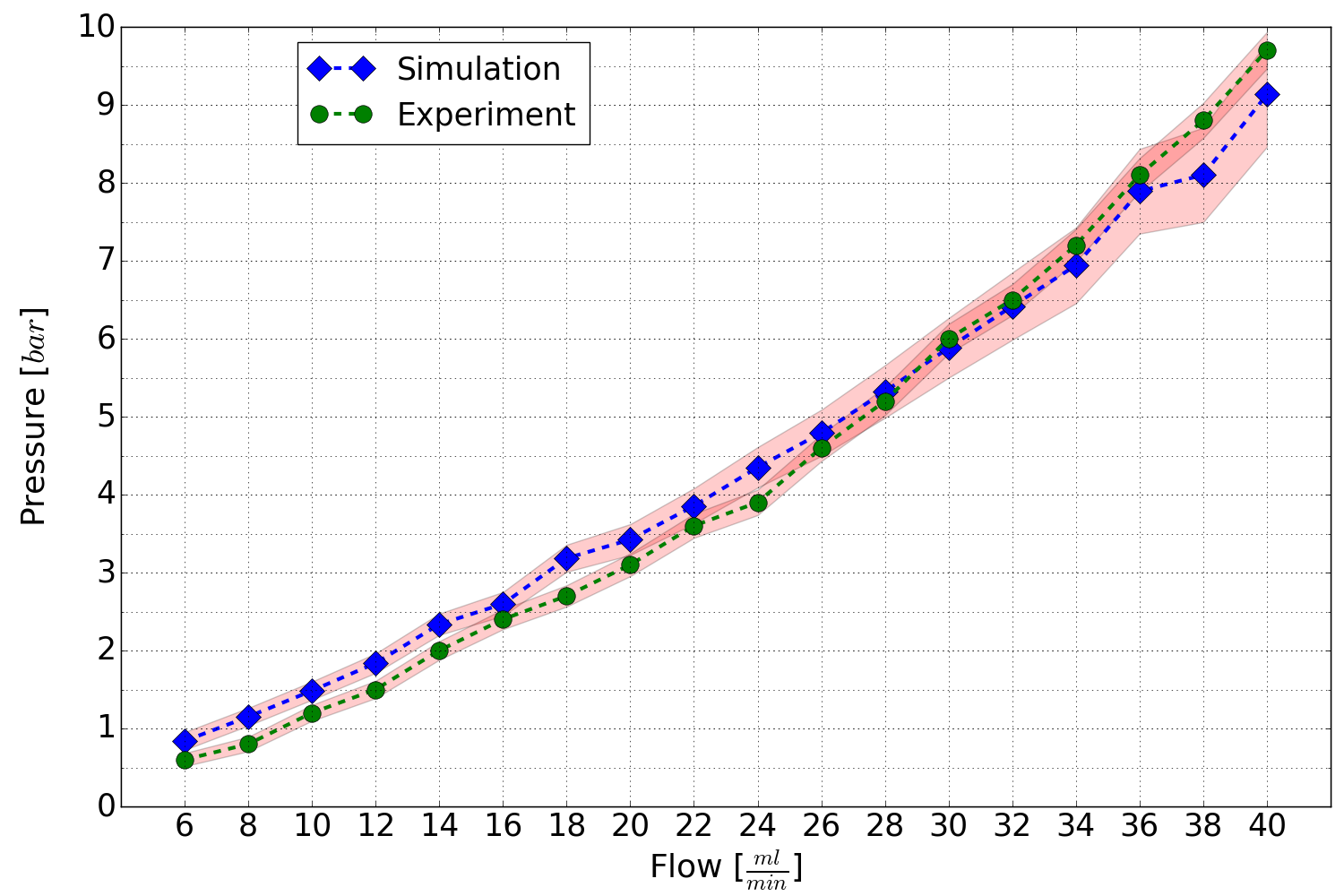}
  \caption{Pressure versus flow comparison between simulation and experiment.}
  \label{fig:pressureVSflow}
\end{figure}
\begin{figure}[tb!]
  \subfloat[]{\includegraphics[width=0.46\textwidth]{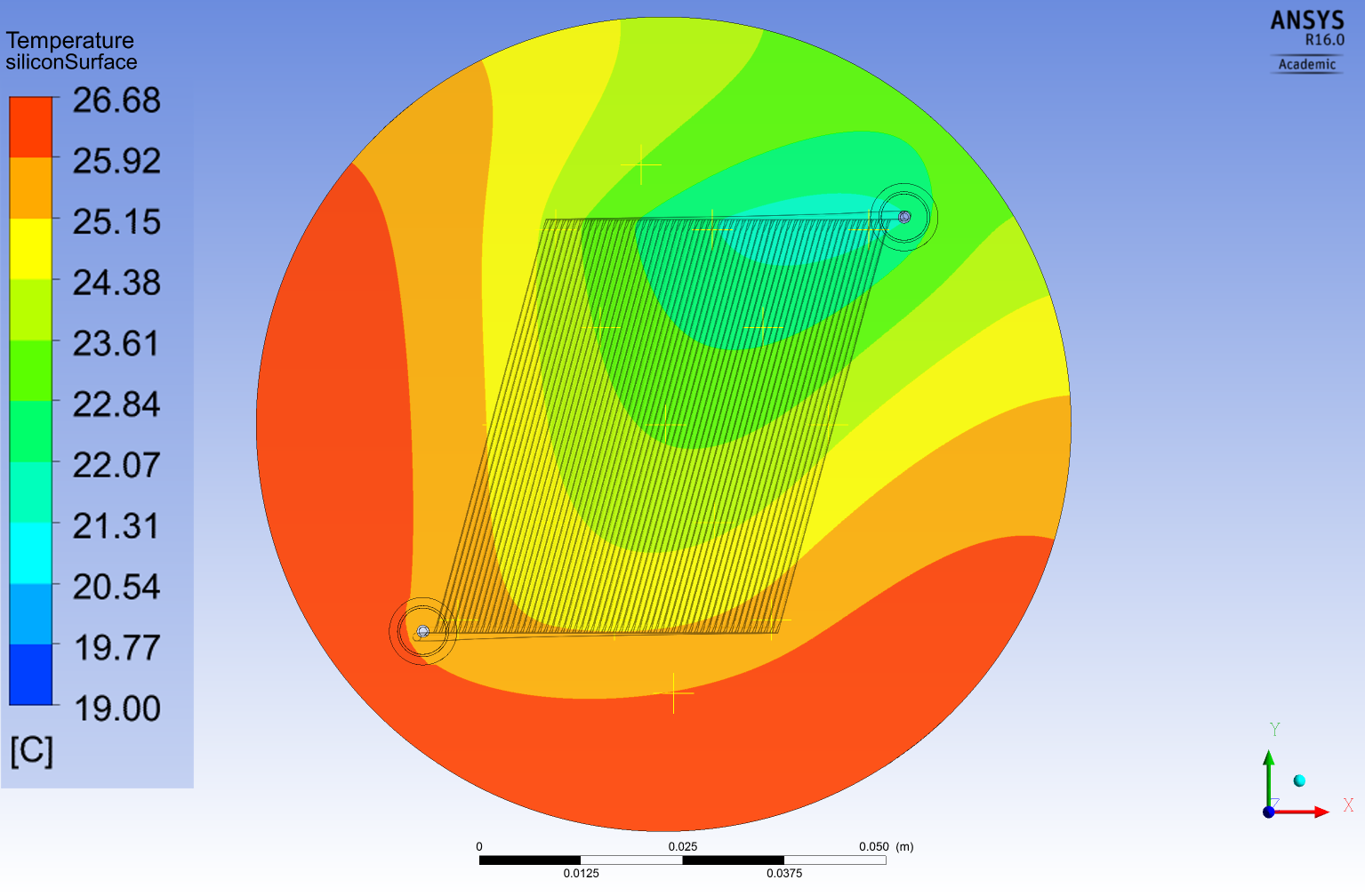}}\\
  \subfloat[]{\includegraphics[width=0.46\textwidth]{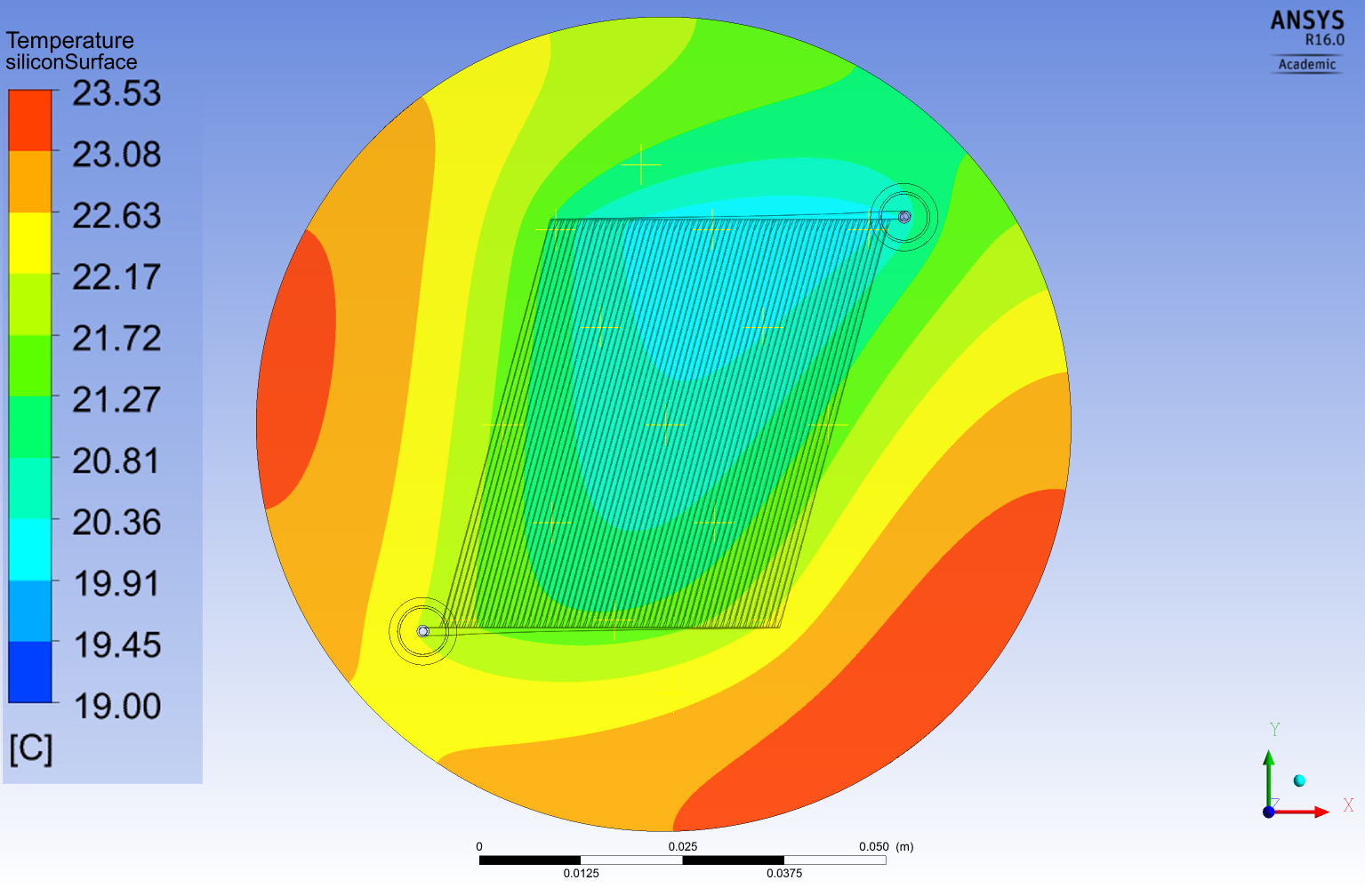}}\\
  \subfloat[]{\includegraphics[width=0.46\textwidth]{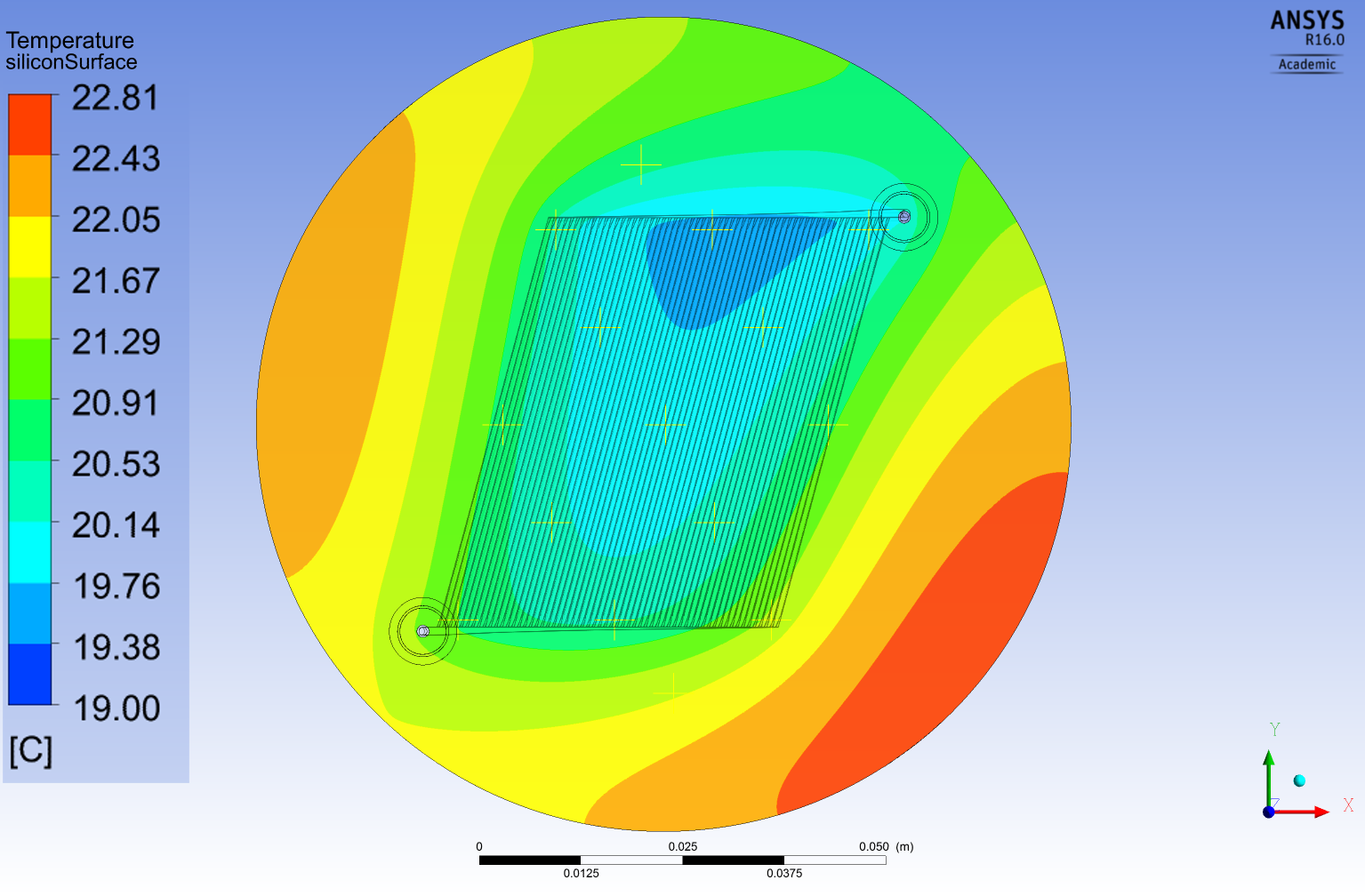}}
  \caption{Simulated temperature maps of the silicon side of the micro-channel wafer, for \SI{30.8}{\mwcm} and the superimposed channel structure. The fluid inlet is at the top right and the outlet is at the bottom left. (a) \SI{10}{\ml\per\minute}, (b) \SI{30}{\ml\per\minute}, (c) \SI{50}{\ml\per\minute}.}
  \label{fig:contour_sim}
\end{figure}
\begin{figure}[tb!]
  \subfloat[]{\includegraphics[width=0.46\textwidth]{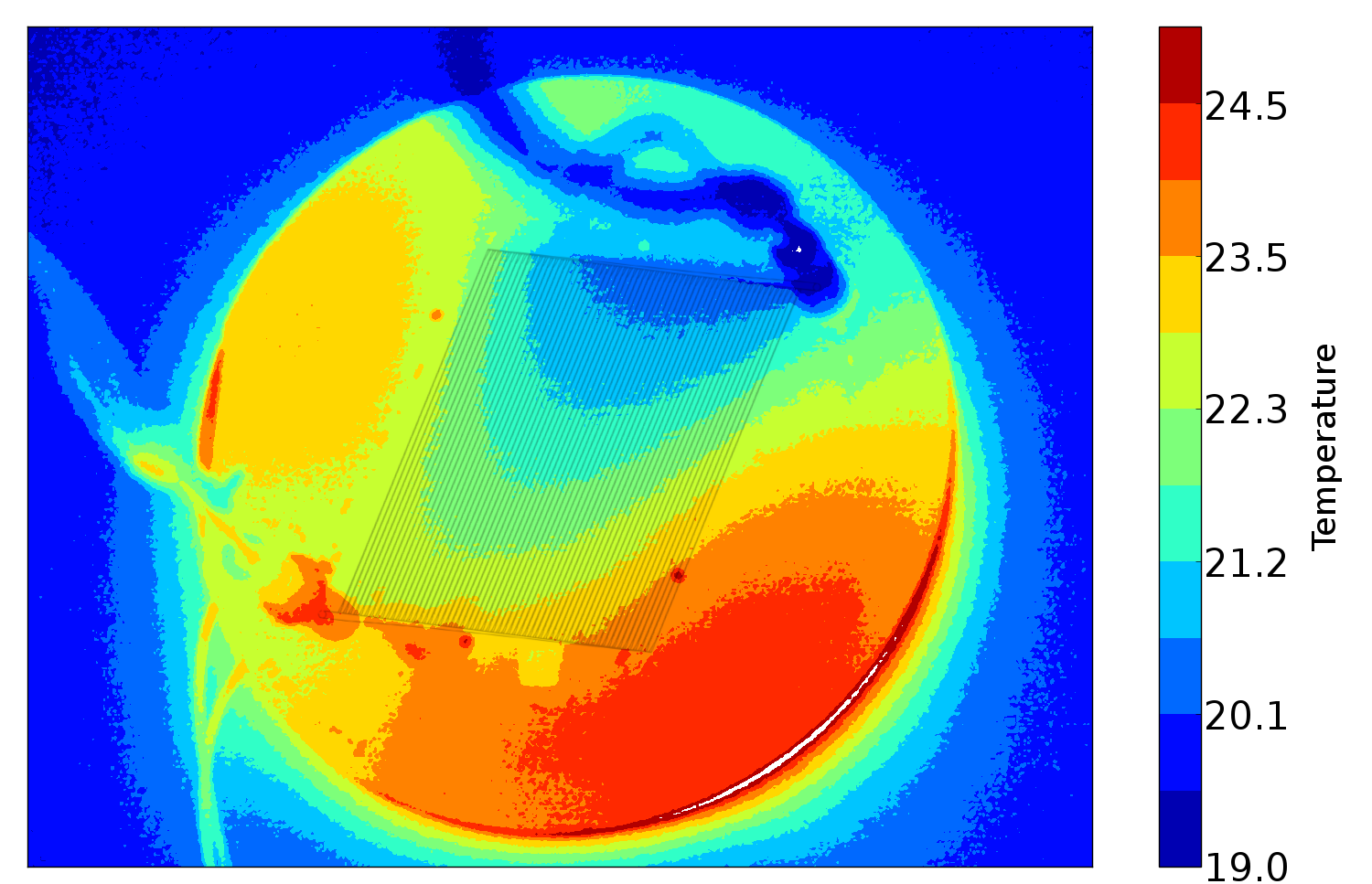}}\\
  \subfloat[]{\includegraphics[width=0.46\textwidth]{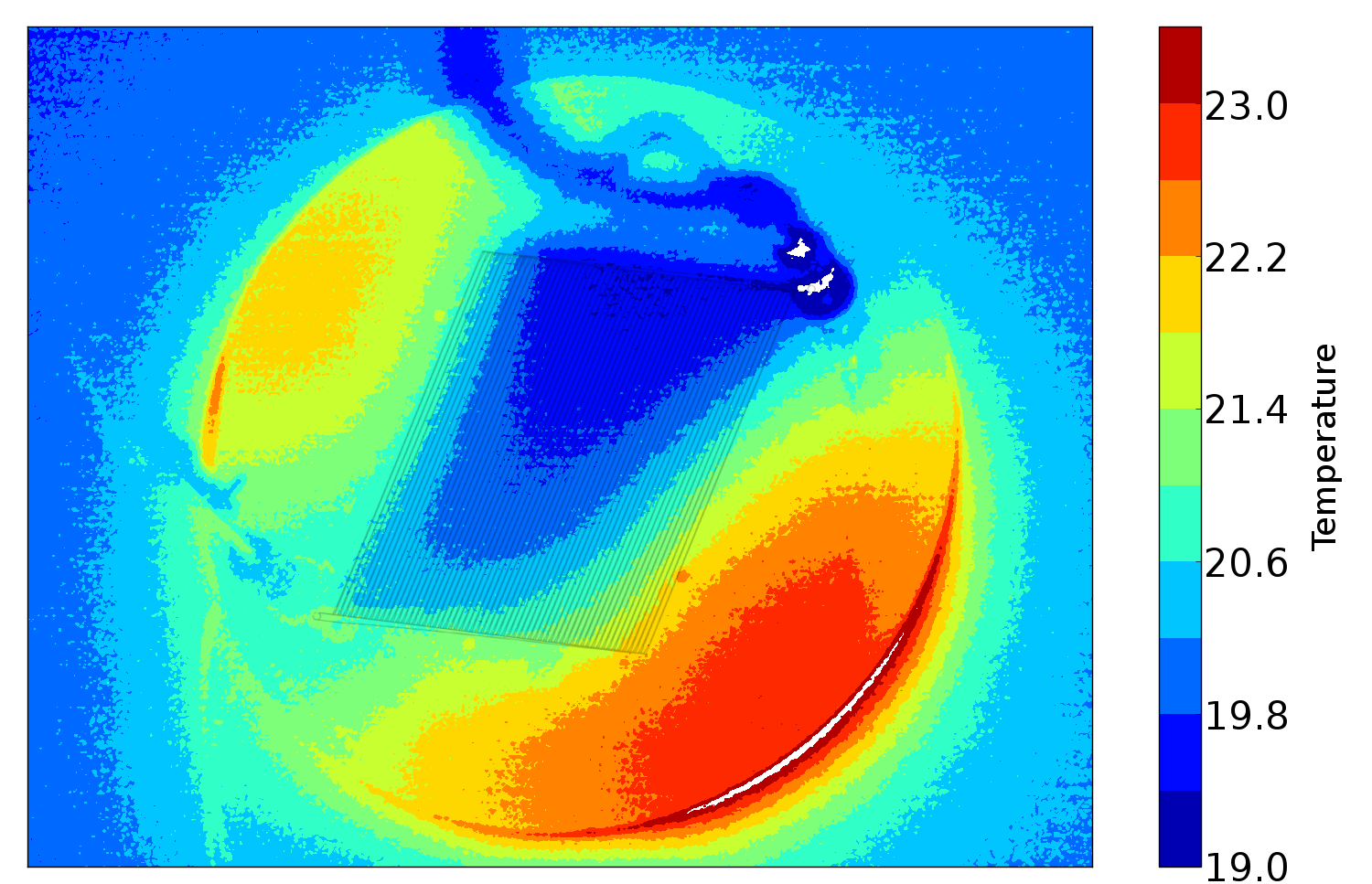}}\\
  \subfloat[]{\includegraphics[width=0.46\textwidth]{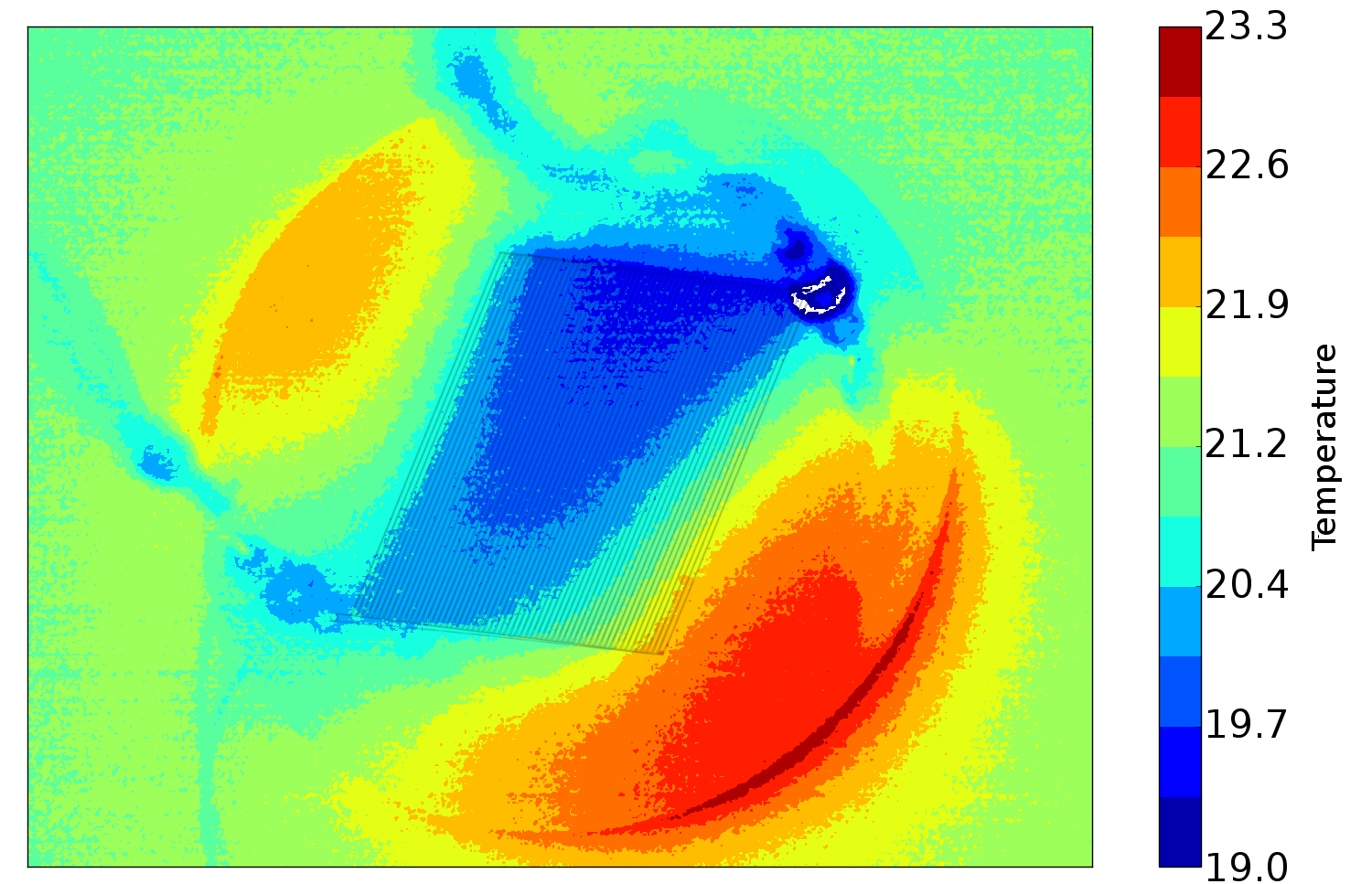}}
  \caption{Temperature maps of the silicon side of the micro-channel wafer taken with an infrared camera with the heater set to \SI{30.8}{\mwcm} and the superimposed channel structure. (a) \SI{10}{\ml\per\minute}, (b) \SI{30}{\ml\per\minute}, (c) \SI{50}{\ml\per\minute}.}
  \label{fig:contour_cam}
\end{figure}
The schematics of the setup are shown in figure \ref{fig:setup_pic}. The setup is built around a high pressure liquid chromatography pump (HPLC) \cite{iota300} aimed at low flow applications providing a maximum flow rate of \SI{300}{\ml\per\minute} and a maximum pressure of \SI{70}{\bar}. A heat exchanger cooled by a cooling unit \cite{huber} is used to control the temperature of the coolant in the main circuit. To accurately measure flow characteristics like flow rate, temperature and density, a flow meter \cite{cori} using the Coriolis method was set right behind the heat exchanger. Two pressure sensors \cite{swage} were used, one before and one after the micro-channel device, to measure the pressure drop through the channel array. The valves \cite{valve} and filters \cite{filter} used do not contribute noticeably to the measured pressure drop. All devices are connected via steel pipes with \US{1/8}{\inch} outer diameter which are converted to steel pipes with \US{1/16}{\inch} outer diameter to connect to the micro-channel inlet and outlet via NanoPort connectors \cite{nanoports}. The following measurements presented in this paper were carried out with the cooling unit cooling the fluid to approximately \SI{19}{\degreeCelsius} at the inlet of the micro-channels and an environment temperature of around \SI{21}{\degreeCelsius}, and the wafer was not insulated or put in any controlled atmosphere.\\
The most crucial measurements of thermal characteristics of a heat sink depend on the temperature difference of the coolant between the inlet and outlet. This value represents directly the heat transported out of the system by the coolant. For that reason two Pt100 temperature dependent resistors are placed on the pipes just at the entry and exit points of the micro-channel device.\\
To be flexible in terms of temperature range a hydrofluoroether \cite{hfe} was chosen as coolant due to its low viscosity at low temperatures (around \SI{0.58}{\gram\per\metre\per\second} at \SI{20}{\degreeCelsius}).\\
For determining the thermal properties of the micro-channel device, the heat load was chosen to satisfy two main criteria. It provided a homogeneous heating and covered the whole wafer surface. For this purpose a \US{4}{\inch} heater \cite{rs} was used, which has exactly the size of the full micro-channel wafer. To measure the temperature on the wafer surface, fifteen Pt100 resistors were placed on the silicon as shown in figure \ref{fig:sensor_order}. In addition, an infrared camera \cite{variocam} was used to map the temperature distribution across the whole wafer surface.\\
In a real detector the power will not be homogeneously distributed over the area to be cooled, but mainly across the readout electronics. This work presents a generic approach to test micro-channels as a cooling structure to transport heat out of a system and since the channel layout is not adapted to a specific detector design, the heat was distributed homogeneously across the wafer.\\
The heater was set to \SI{30.8}{\mwcm}, \SI{43.1}{\mwcm}, \SI{55.5}{\mwcm} and \SI{67.8}{\mwcm}, which corresponds to \SI{2.5}{\W}, \SI{3.5}{\W}, \SI{4.5}{\W} and \SI{5.5}{\W} over the full wafer surface. Running the setup at different heat loads can reveal the weak points of the layout or material composition. The following measurements were performed at an inlet temperature of around \SI{19}{\degreeCelsius} and flow rates ranged from \SIrange{10}{50}{\ml\per\minute} in \SI{5}{\ml\per\minute} steps.
%%%%%%%%%%%%%%%%%%%%%%%%%%%%%%%%%%%%%%%%%%%%%%%%%%%%%%%%%%%%%%%%
%%%%%%%%%%%%%%%%%%%%%%%%%%%%%%%%%%%%%%%%%%%%%%%%%%%%%%%%%%%%%%%%
%%%%%%%%%%%%%%%%%%%%%%%%%%%%%%%%%%%%%%%%%%%%%%%%%%%%%%%%%%%%%%%%

\subsection{Experiment results}
\label{exp_res}
\begin{figure}[tb]
  \includegraphics[width=0.46\textwidth]{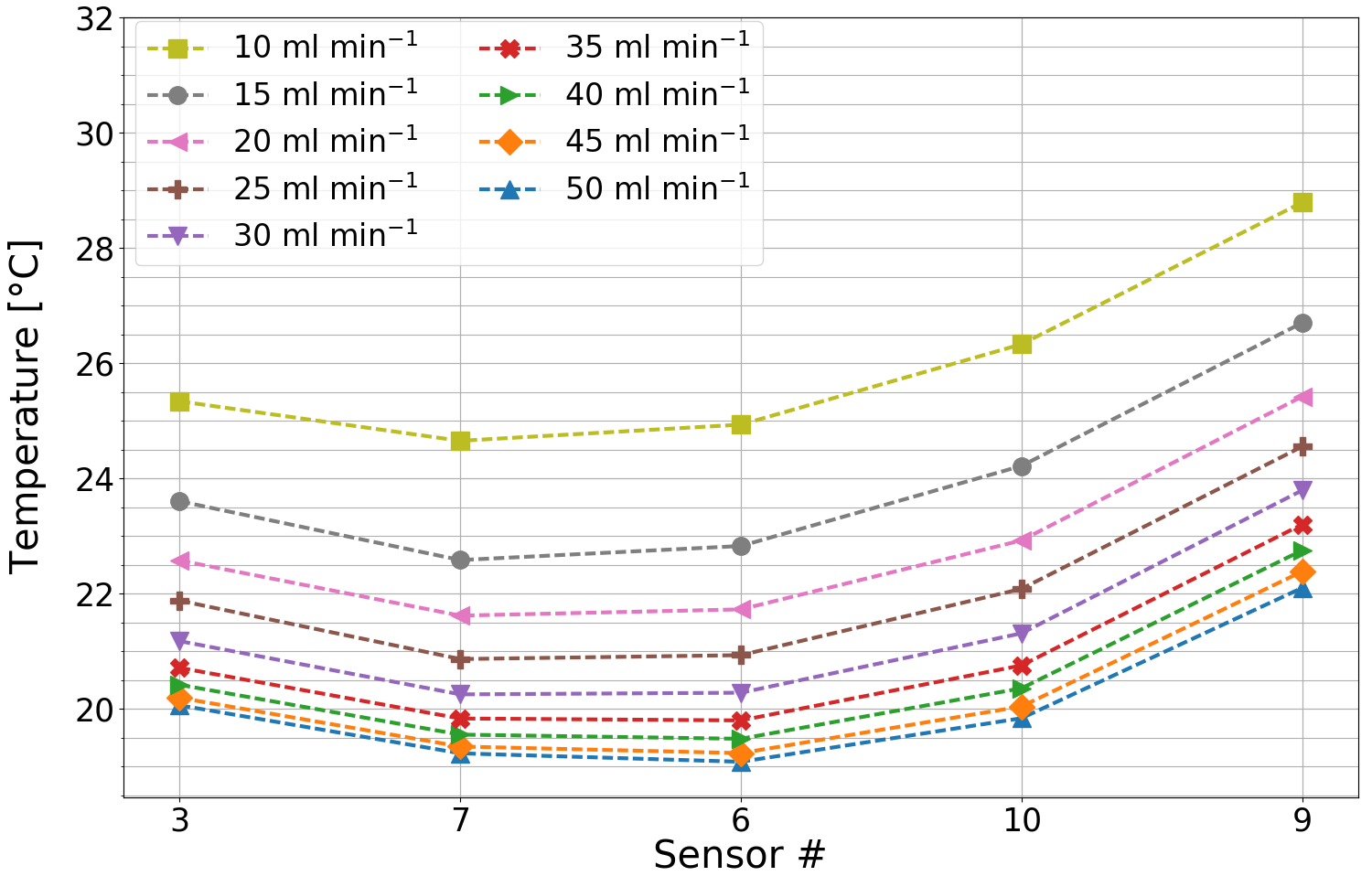}
  \caption{Temperature measured along line 1, as indicated in figure \ref{fig:sensor_order}.}
  \label{fig:diagonals1}
\end{figure}

\begin{figure}[tb]
  \includegraphics[width=0.46\textwidth]{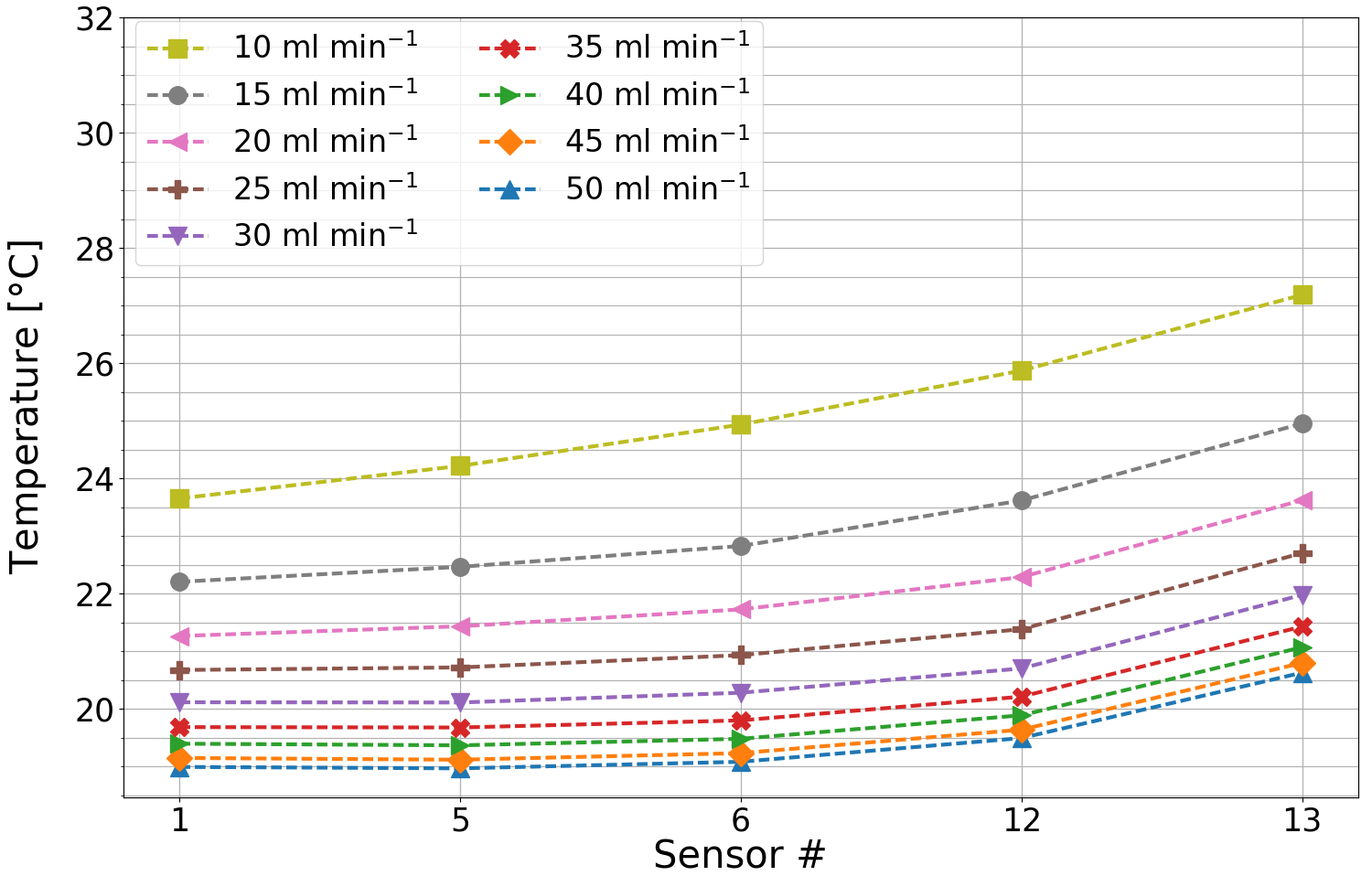}
  \caption{Temperature measured along line 2, as indicated in figure \ref{fig:sensor_order}.}
  \label{fig:diagonals2}
\end{figure}

\begin{figure}[tb]
  \includegraphics[width=0.46\textwidth]{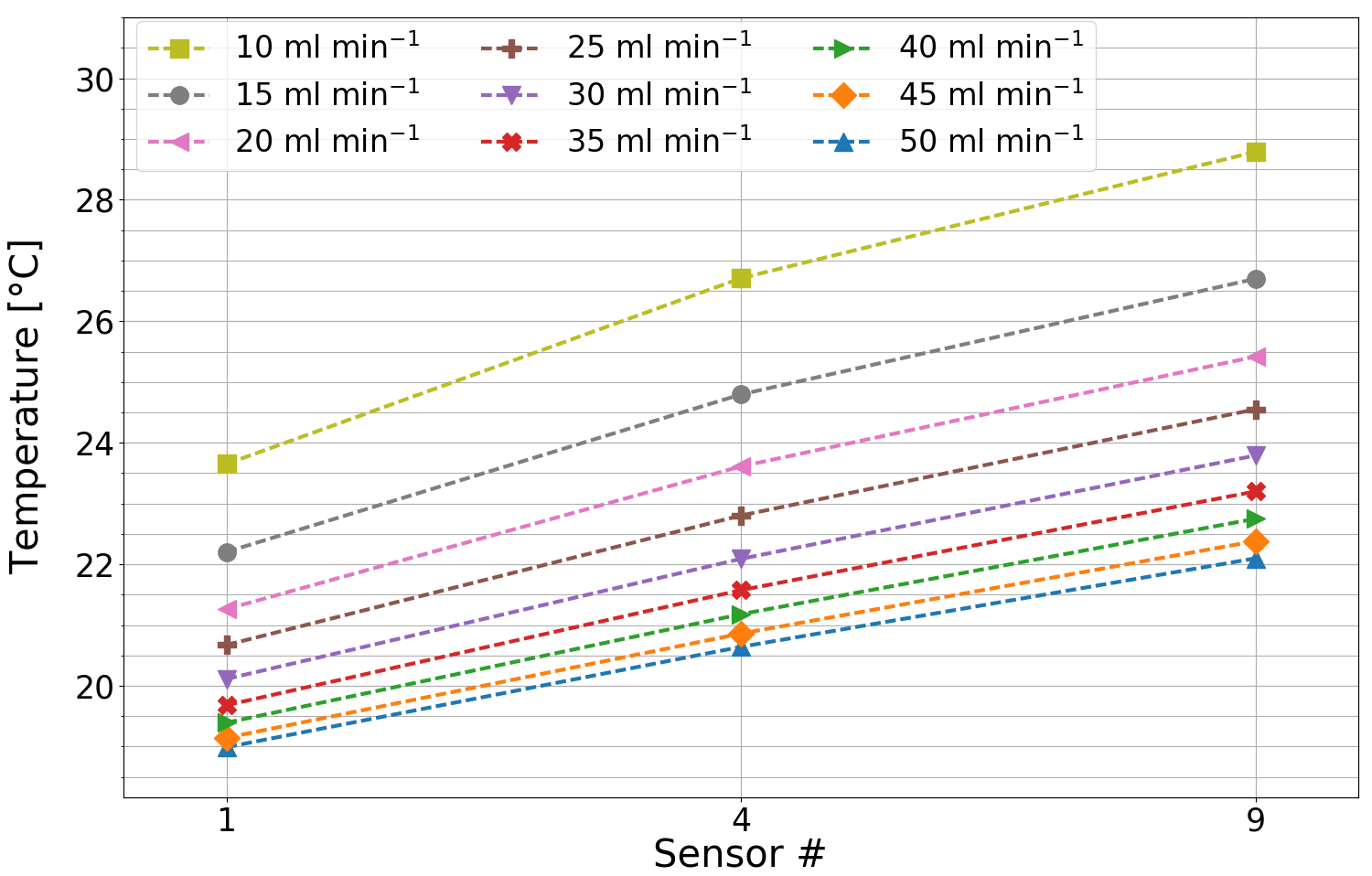}
  \caption{Temperature measured along line 3, as indicated in figure \ref{fig:sensor_order}.}
  \label{fig:lines1}
\end{figure}

\begin{figure}[tb]
  \includegraphics[width=0.46\textwidth]{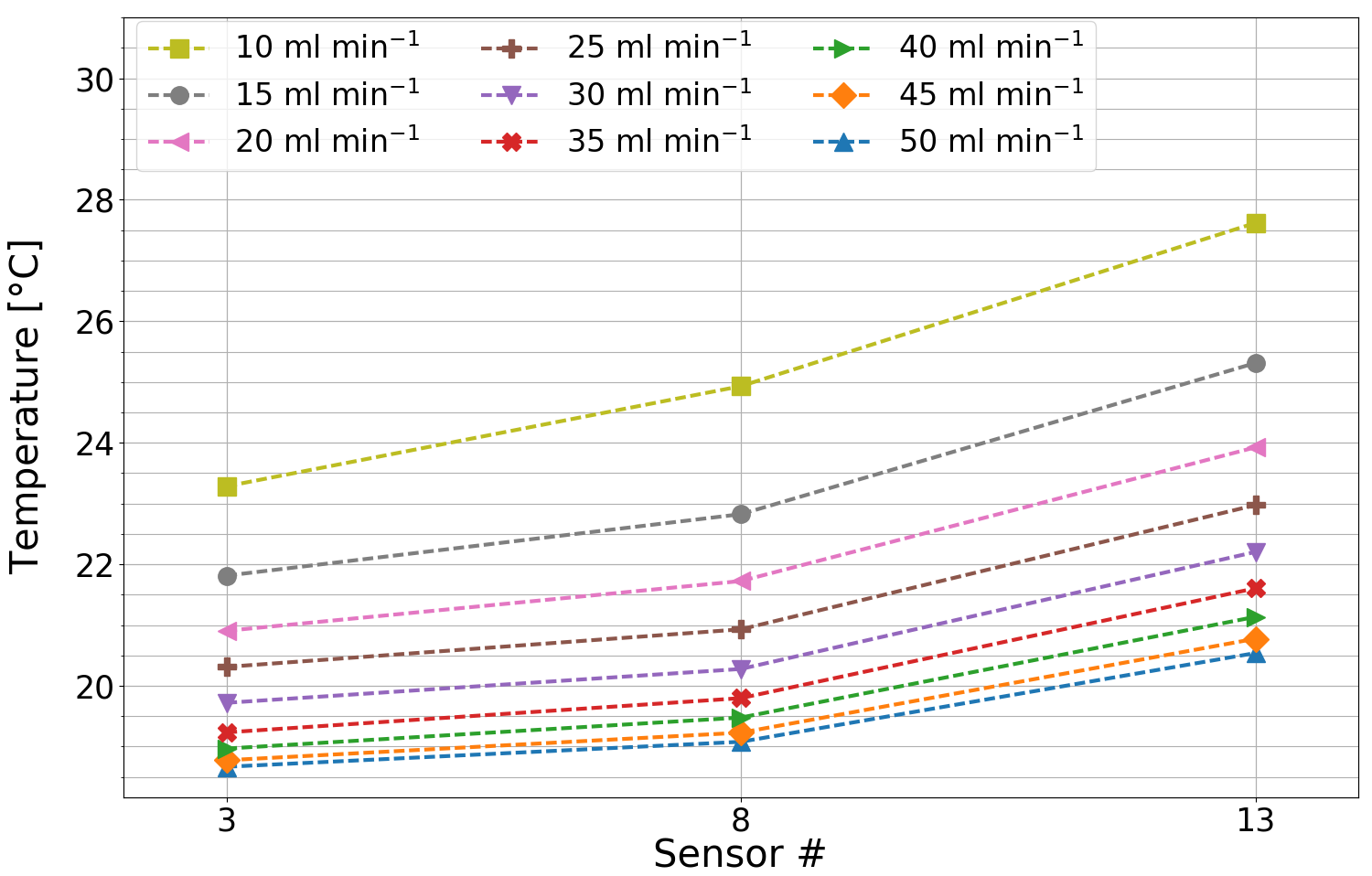}
  \caption{Temperature measured along line 4, as indicated in figure \ref{fig:sensor_order}.}
  \label{fig:lines5}
\end{figure}

\begin{figure}[htb]
  \includegraphics[width=0.46\textwidth]{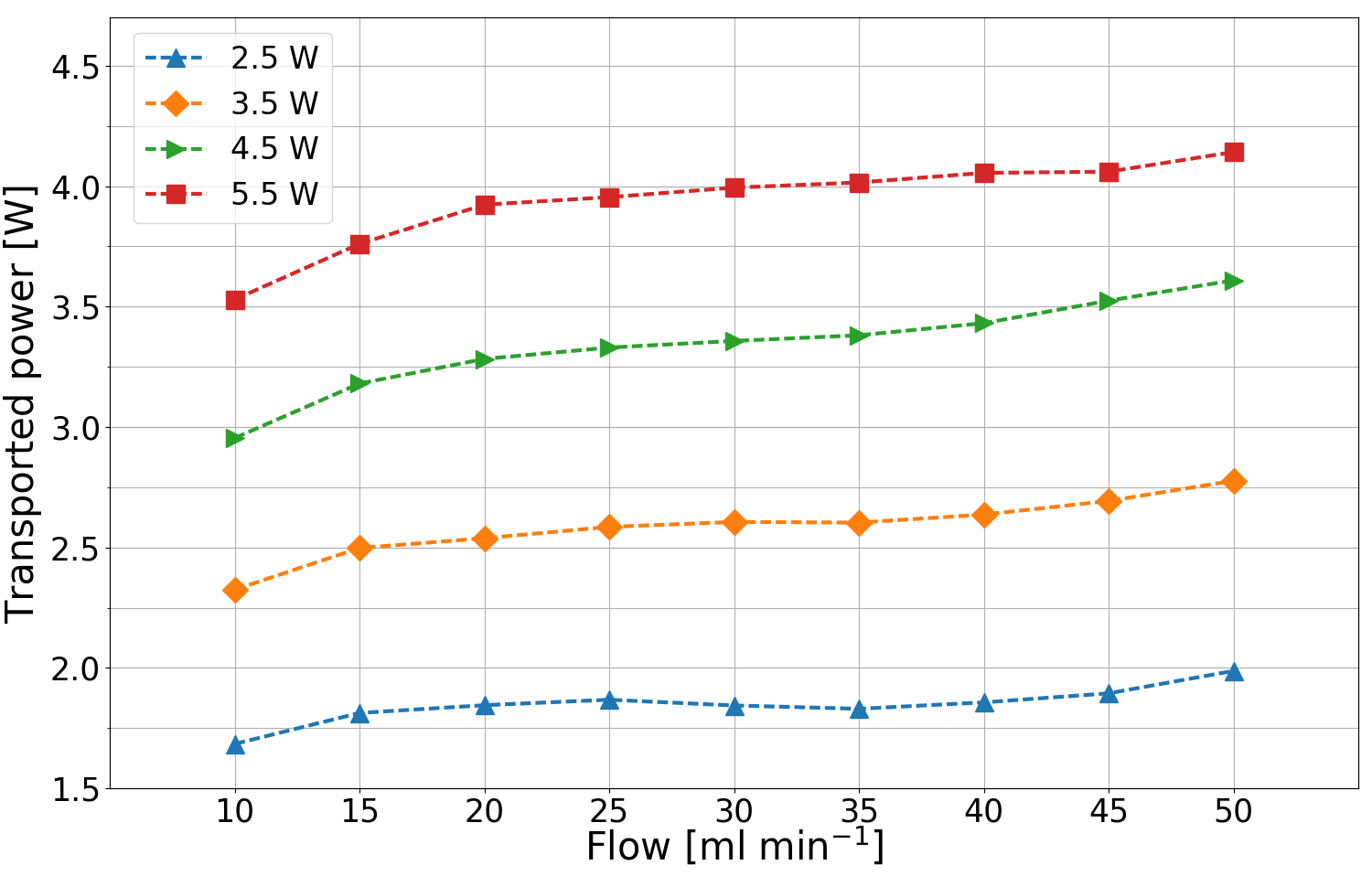}
  \caption{Heat transported by the coolant for different flow rates and heater settings.}
  \label{fig:evacuated}
\end{figure}

\begin{figure}[tb]
  \includegraphics[width=0.46\textwidth]{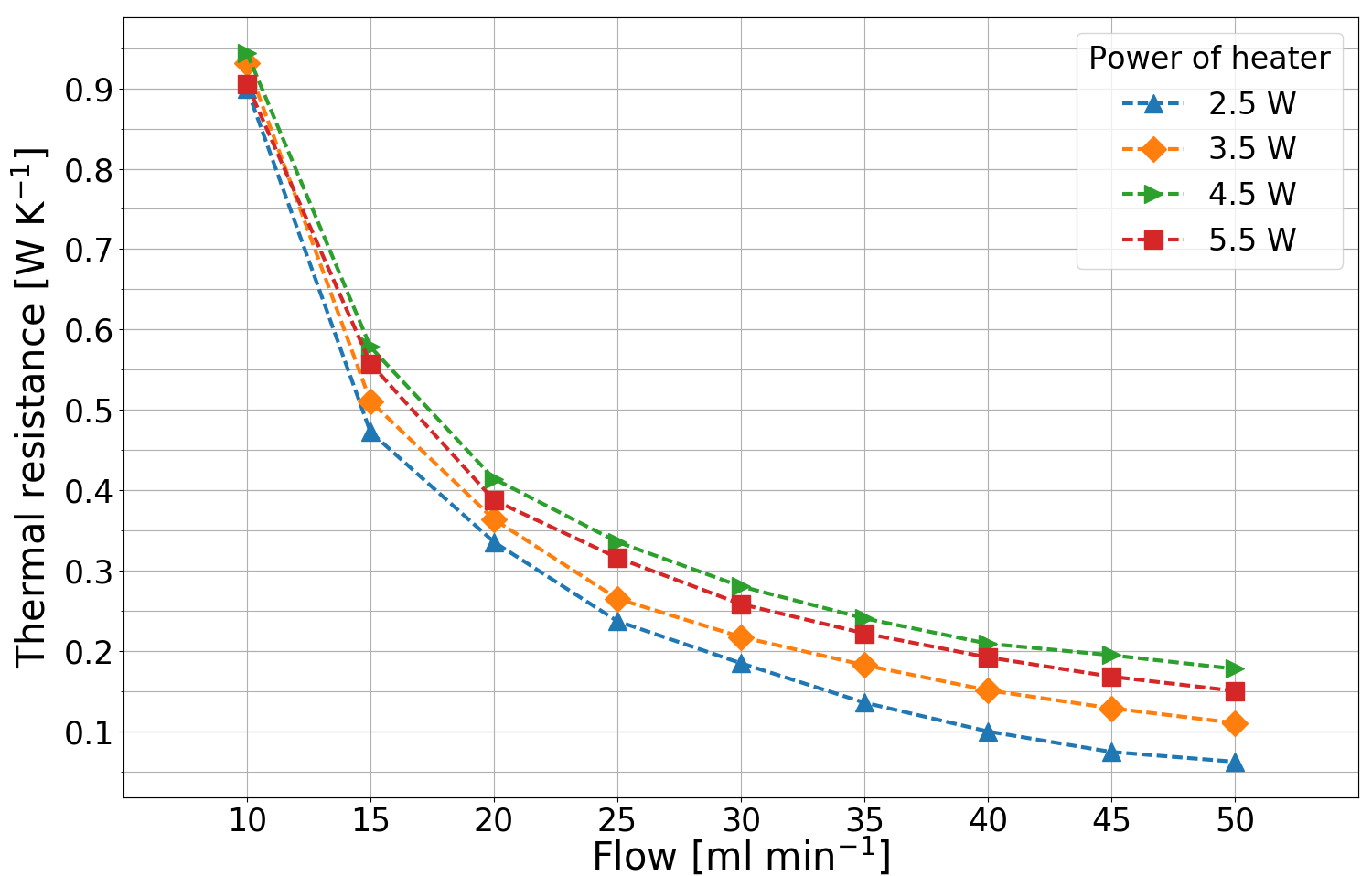}
  \caption{Measured thermal resistance as a function of flow rate. The divergence of the measured values towards higher flow rates is caused by the low thermal conductivity of the Pyrex.}
  \label{fig:thermResistance}
\end{figure}

Figure \ref{fig:pressureVSflow} compares the fluid pressure at the inlet of the micro-channel structure versus the total flow through the device for both the simulations and the experiment. A possible explanation for the different slopes observed in the simulation and the measurement crossing each other at around \SI{34}{\ml\per\minute} is the increased wall surface area, due to the scalloping mentioned in section \ref{Fabrication}. The increased surface area of the channel walls might increase the pressure drop at higher flow rates, which is missing in the simulation. Nevertheless, the simulated values are in a good overall agreement with the experiment. The mesh in the simulation was refined until the results did not improve anymore, and the only differences were fluctuations in the obtained pressure drop values. The fluctuations increased with higher flow rates and the error on the simulated data in figure \ref{fig:pressureVSflow} is based on these remaining fluctuations. The pulsations of the pump are causing a deviation of the desired flow rate. The uncertainty due to the precision of the pressure sensor is negligible in comparison to the error caused by the pulsations. The deviation from the preset flow rate is represented by the red error band surrounding the experimental measurement data points.\\
Destructive tests were performed, showing that the wafer broke near the inlet at a critical pressure of over \SI{30}{\bar}.\\
The temperature measurements obtained with the infrared camera were used primarily to study the temperature distribution across the wafer, whereas the Pt100 resistors were used to obtain precise measurements at certain points on the wafer. An offset measurement of all resistors at a fixed temperature was used to calibrate them to one reference resistor. The results of the measurements with the infrared camera can be seen in figure \ref{fig:contour_cam}. The temperature maps in figure \ref{fig:contour_cam}, as well as the simulation results in figure \ref{fig:contour_sim}, both corresponding to a heat load of \SI{2.5}{\W} or a flux of \SI{30.8}{\mwcm}, show that the micro-channel structure has a tendency to cool the left side (outlet) more than the right (inlet). To visualize the temperature distribution across the wafer, the temperature measured at four lines connecting measurement-points are plotted at different flow rates (see figure \ref{fig:sensor_order}).\\
The two diagonals (line 1 and 2) in figure \ref{fig:diagonals1} and \ref{fig:diagonals2} are along the fluid direction but are not following one single channel. Line 1 has a maximal temperature difference of \SI{3.02}{\degreeCelsius} at the maximum flow rate and line 2 with a maximal temperature difference of \SI{1.66}{\degreeCelsius}. The two diagonals contain points affected by different layout properties. One contains both inlet and outlet whereas the other one contains the Pt100 resistors at the outer corners of the layout with less flow.\\
In figures \ref{fig:lines1} and \ref{fig:lines5} the temperature along the lines 3 and 4 - as indicated in figure \ref{fig:sensor_order} - are plotted. Instead of possessing similar temperature gradients, on line 3 the maximal difference of \SI{2.4}{\degreeCelsius} is much higher than the one on line 4 with a difference of \SI{0.7}{\degreeCelsius}. This is due to the effect of the cold fluent entering at Pt100 resistor 1 and heating up on the way to Pt100 resistor 3.\\
In total the maximum temperature difference across the wafer surface is \SI{3.4}{\degreeCelsius} at a flow rate of \SI{50}{\ml\per\minute} and a heat load of \SI{30.8}{\mwcm}.\\
As the difference between the inlet and the outlet temperature directly reflects the heat $q_{r}$ transported out of the system, the measurement of the temperature at the inlet and the outlet and the coolant specifications can be used to obtain
\begin{equation}
  \label{eq:evacuated}
  q_{r} = \dot m C_{p} \Delta T,
\end{equation}
with $\dot m$ the mass flow, $C_{p}$ the specific heat capacity and $\Delta T$ the temperature difference of the coolant between inlet and outlet. Figure \ref{fig:evacuated} shows the heat removed by the coolant for different flow rates. As can be seen, the values differ from the heater settings by at least 20\%. The difference between $q_{r}$ and the preset heater power represents the losses due to free convection and radiation to the surrounding air and conduction to the Pt100 resistors and NanoPort connectors.\\
The behavior of the effects can also be visualized by a common way to characterize heat sinks, namely by its thermal resistance
\begin{equation}
  \label{eq:thermalResistance}
  R = \frac{\overline{T_{s}} - \overline{T_{f}}}{q_{r}},
\end{equation}
which in this case is the difference between the mean fluid temperature $\overline{T_{f}}$ and the mean surface temperature $\overline{T_{s}}$ for the transported heat $q_{r}$. The mean surface temperature is defined only by the temperatures measured in the middle of the wafer with the Pt100 resistors 5, 6, 7, 10 and 12, to suppress the influence of edge effects - caused by a surplus of not actively cooled material - at the outer border of the channel structure. The thermal resistance, as shown in figure \ref{fig:thermResistance}, is decreasing with increasing flow rate. A saturation effect is visible for higher flow rates. The difference between the thermal resistance for different heat loads is increasing with higher flow rates, which is caused by the low thermal conductivity of the Pyrex causing a large uncertainty in the measurement of the thermal resistance. 

%%%%%%%%%%%%%%%%%%%%%%%%%%%%%%%%%%%%%%%%%%%%%%%%%%%%%%%%%%%%%%%%
%%%%%%%%%%%%%%%%%%%%%%%%%%%%%%%%%%%%%%%%%%%%%%%%%%%%%%%%%%%%%%%%
%%%%%%%%%%%%%%%%%%%%%%%%%%%%%%%%%%%%%%%%%%%%%%%%%%%%%%%%%%%%%%%%

\subsection{Comparison with simulation}
\label{comparison}
\begin{figure}[tb]
  \includegraphics[width=0.46\textwidth]{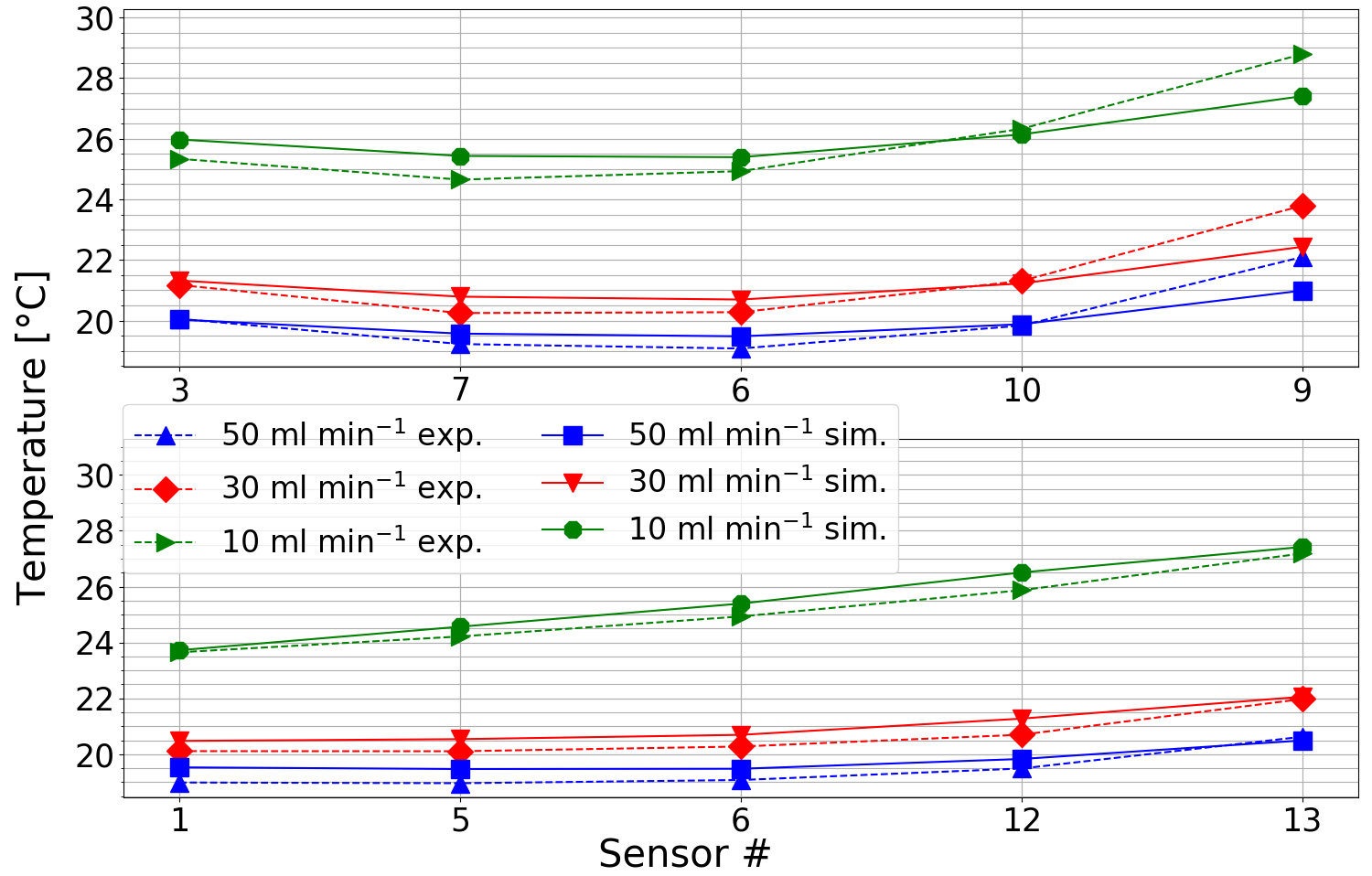}
  \caption{The temperature of the Pt100 resistors on the lines 1 and 2 in the simulation and in the experiment compared with each other for a heat load of \SI{30.8}{\mwcm}.}
  \label{fig:diagonalCompare}
\end{figure}

The temperature distribution along the wafer can already be compared through the results from the simulation in figure \ref{fig:contour_sim} with the infrared images in figure \ref{fig:contour_cam}.\\
There is a good agreement between simulation and experimental results, although some slight differences are visible in the areas around the outlet. A more detailed comparison of individual points on the wafer can be seen in figure \ref{fig:diagonalCompare}.
The inlet temperature was fixed in the simulation and was slightly varying in the experiment due to environmental influences. This discrepancy has been taken into account in figure \ref{fig:diagonalCompare}. The measurements were also affected by small variations of the ambient temperature. However, the influence of the ambient air was not implemented in the simulation.

%%%%%%%%%%%%%%%%%%%%%%%%%%%%%%%%%%%%%%%%%%%%%%%%%%%%%%%%%%%%%%%%
%%%%%%%%%%%%%%%%%%%%%%%%%%%%%%%%%%%%%%%%%%%%%%%%%%%%%%%%%%%%%%%%
%%%%%%%%%%%%%%%%%%%%%%%%%%%%%%%%%%%%%%%%%%%%%%%%%%%%%%%%%%%%%%%%

\section{Conclusion}
\label{conclusion}
A micro-channel structure was designed and produced by DRIE and anodic bonding. The pressure at the inlet reached a maximum of around \SI{15}{\bar} at flow rates around \SI{50}{\ml\per\minute}, which does not put the device at risk of breaking. Volumetric flows varying from few \si{\ml\per\minute} to \SI{50}{\ml\per\minute} were tested and found to be able to remove enough heat from the micro-channel device to keep the maximum temperature difference over the micro-channel structure below \SI{4}{\degreeCelsius} for a heat load of \SI{30.8}{\mwcm}.\\
Simulations were set up and tuned with the results of the experiment leading to similar results. An additional tuning of the simulation could be carried out using the values of the actual heat removed by the fluid in the experiment shown in figure \ref{fig:evacuated} to have an accurate model of the heat transfer to the environment within the simulation. Further studies have to address the effect of the remaining material surrounding the micro-channel structure on the thermal measurements and deviations between measurements and the simulations.\\
Flow and thermal tests show its general feasibility as a large area cooling device. This study demonstrates the first steps of a generic approach of developing a micro-channel structure from silicon. A manifold shape was developed, providing a homogeneous flow across a large channel array, with the potential of being adopted to other sizes, which has not yet been addressed by other HEP experiments cited in this work.

%%%%%%%%%%%%%%%%%%%%%%%%%%%%%%%%%%%%%%%%%%%%%%%%%%%%%%%%%%%%%%%%
%%%%%%%%%%%%%%%%%%%%%%%%%%%%%%%%%%%%%%%%%%%%%%%%%%%%%%%%%%%%%%%%
%%%%%%%%%%%%%%%%%%%%%%%%%%%%%%%%%%%%%%%%%%%%%%%%%%%%%%%%%%%%%%%%

\section{Acknowledgment}
\label{acknowledgment}
The authors want to thank the colleagues from the electronic and mechanical workshop at DESY and especially Torsten K\"ulper and Uwe Packheiser for their support. Martin Lemke from the Service-Zentrum Mechanik at DESY was a great help for the work on simulating the micro-channel device.\\

%% The Appendices part is started with the command \appendix; appendix sections are then done as normal sections


%% References
%%
%% Following citation commands can be used in the body text: Usage of \cite is as follows: \cite{key} ==>> [#] \cite[chap. 2]{key} ==>> [#, chap. 2] \citet{key} ==>> Author [#]

\raggedright \bibliographystyle{model1a-num-names} \bibliography{mcc_database.bib}

\begin{thebibliography}{24}
\expandafter\ifx\csname natexlab\endcsname\relax\def\natexlab#1{#1}\fi
\providecommand{\bibinfo}[2]{#2}
\ifx\xfnm\relax \def\xfnm[#1]{\unskip,\space#1}\fi
%Type = Article
\bibitem[{Tuckerman and Pease(1981)}]{Tuckerman1981}
\bibinfo{author}{D.~Tuckerman}, \bibinfo{author}{R.~Pease},
\newblock \bibinfo{title}{{High-performance heat sinking for VLSI}},
\newblock \bibinfo{journal}{IEEE Electron Device Letters} \bibinfo{volume}{2}
  (\bibinfo{year}{1981}) \bibinfo{pages}{126--129}.
%Type = Article
\bibitem[{Sakanova et~al.(2015)Sakanova, Keian, and Zhao}]{AsselSakanovaa}
\bibinfo{author}{A.~Sakanova}, \bibinfo{author}{C.~C. Keian},
  \bibinfo{author}{J.~Zhao},
\newblock \bibinfo{title}{{Performance improvements of microchannel heat sink
  using wavy channel and nanofluids}},
\newblock \bibinfo{journal}{International Journal of Heat and Mass Transfer}
  \bibinfo{volume}{89} (\bibinfo{year}{2015}) \bibinfo{pages}{59--74}.
%Type = Article
\bibitem[{Attree et~al.(2008)Attree, Anderson, Anderssen, Akhnazarov, and
  Bates}]{Search2008}
\bibinfo{author}{D.~Attree}, \bibinfo{author}{B.~Anderson},
  \bibinfo{author}{E.~C. Anderssen}, \bibinfo{author}{V.~Akhnazarov},
  \bibinfo{author}{R.~L. Bates},
\newblock \bibinfo{title}{{The evaporative cooling system for the ATLAS inner
  detector}},
\newblock \bibinfo{journal}{Journal of Instrumentation} \bibinfo{volume}{07003}
  (\bibinfo{year}{2008}) \bibinfo{pages}{1--36}.
%Type = Misc
\bibitem[{hfe(2016)}]{hfe}
\bibinfo{title}{{3M HFE 7100}}, \bibinfo{howpublished}{{http://www.3m.com}},
  \bibinfo{year}{2016}.
%Type = Article
\bibitem[{Francescon et~al.(2015)Francescon, Romagnoli, Mapelli, Petagna,
  Gargiulo, Musa, Thome, and {Del Col}}]{alice}
\bibinfo{author}{A.~Francescon}, \bibinfo{author}{G.~Romagnoli},
  \bibinfo{author}{A.~Mapelli}, \bibinfo{author}{P.~Petagna},
  \bibinfo{author}{C.~Gargiulo}, \bibinfo{author}{L.~Musa},
  \bibinfo{author}{J.~R. Thome}, \bibinfo{author}{D.~{Del Col}},
\newblock \bibinfo{title}{{Development of interconnected silicon
  micro-evaporators for the on-detector electronics cooling of the future ITS
  detector in the ALICE experiment at LHC}},
\newblock \bibinfo{journal}{Applied Thermal Engineering} \bibinfo{volume}{93}
  (\bibinfo{year}{2015}) \bibinfo{pages}{1367--1376}.
%Type = Article
\bibitem[{Romagnoli et~al.(2015)Romagnoli, Feito, Brunel, Catinaccio, Degrange,
  Mapelli, Morel, Noel, and Petagna}]{Romagnoli2015}
\bibinfo{author}{G.~Romagnoli}, \bibinfo{author}{D.~A. Feito},
  \bibinfo{author}{B.~Brunel}, \bibinfo{author}{A.~Catinaccio},
  \bibinfo{author}{J.~Degrange}, \bibinfo{author}{A.~Mapelli},
  \bibinfo{author}{M.~Morel}, \bibinfo{author}{J.~Noel},
  \bibinfo{author}{P.~Petagna},
\newblock \bibinfo{title}{{Silicon micro-fluidic cooling for NA62 GTK pixel
  detectors}},
\newblock \bibinfo{journal}{Microelectronic Engineering} \bibinfo{volume}{145}
  (\bibinfo{year}{2015}) \bibinfo{pages}{133--137}.
%Type = Article
\bibitem[{Nomerotski et~al.(2013)Nomerotski, Buytart, Collins, Dumps, Greening,
  John, Mapelli, Leflat, Li, Romagnoli, and Verlaat}]{Nomerotski}
\bibinfo{author}{A.~Nomerotski}, \bibinfo{author}{J.~Buytart},
  \bibinfo{author}{P.~Collins}, \bibinfo{author}{R.~Dumps},
  \bibinfo{author}{E.~Greening}, \bibinfo{author}{M.~John},
  \bibinfo{author}{A.~Mapelli}, \bibinfo{author}{A.~Leflat},
  \bibinfo{author}{Y.~Li}, \bibinfo{author}{G.~Romagnoli},
  \bibinfo{author}{B.~Verlaat},
\newblock \bibinfo{title}{{Evaporative CO$_{2}$ cooling using microchannels
  etched in silicon for the future LHCb vertex detector Module prototype with
  microchannel cooling}},
\newblock \bibinfo{journal}{Journal of Instrumentation}  (\bibinfo{year}{2013})
  \bibinfo{pages}{1--12}.
%Type = Book
\bibitem[{Munson et~al.(2001)Munson, Young, and Okiishi}]{fluidFundamentals}
\bibinfo{author}{B.~R. Munson}, \bibinfo{author}{D.~F. Young},
  \bibinfo{author}{T.~H. Okiishi}, \bibinfo{title}{{Fundamentals of fluid
  mechanics}}, volume~\bibinfo{volume}{16}, \bibinfo{publisher}{Wiley},
  \bibinfo{edition}{4th} edition, \bibinfo{year}{2001}.
%Type = Misc
\bibitem[{Laermer and Schilp(1996)}]{laermer}
\bibinfo{author}{F.~Laermer}, \bibinfo{author}{A.~Schilp},
  \bibinfo{title}{Method of anisotropically etching silicon},
  \bibinfo{year}{1996}. \bibinfo{note}{US Patent 5,501,893}.
%Type = Article
\bibitem[{Ayon et~al.(1999)Ayon, Braff, Lin, Sawin, and Schmidt}]{plasma}
\bibinfo{author}{A.~A. Ayon}, \bibinfo{author}{R.~Braff},
  \bibinfo{author}{C.~C. Lin}, \bibinfo{author}{H.~Sawin},
  \bibinfo{author}{M.~A. Schmidt},
\newblock \bibinfo{title}{{Characterization of a time multiplexed inductively
  coupled plasma etcher}},
\newblock \bibinfo{journal}{J. Electrochem. Soc.}  (\bibinfo{year}{1999})
  \bibinfo{pages}{pp 339}.
%Type = Book
\bibitem[{Madou(2002)}]{microfabrication}
\bibinfo{author}{M.~J. Madou}, \bibinfo{title}{{Fundamentals of
  microfabrication: the science of miniaturization}}, \bibinfo{publisher}{CRC
  Press}, \bibinfo{year}{2002}. \bibinfo{note}{2nd edition}.
%Type = Article
\bibitem[{Knowles and Van~Helvoort(2006)}]{bonding}
\bibinfo{author}{K.~M. Knowles}, \bibinfo{author}{A.~T.~J. Van~Helvoort},
\newblock \bibinfo{title}{{Anodic bonding}},
\newblock \bibinfo{journal}{International Materials Reviews}
  (\bibinfo{year}{2006}) \bibinfo{pages}{pp. 273--311}.
%Type = Misc
\bibitem[{ope(2016)}]{openfoam}
\bibinfo{title}{{OpenFoam, open source CFD software}},
  \bibinfo{howpublished}{{http://www.openfoam.com/}}, \bibinfo{year}{2016}.
%Type = Misc
\bibitem[{ans(2016)}]{ansys}
\bibinfo{title}{{ANSYS, engineering simulation}},
  \bibinfo{howpublished}{{http://www.ansys.com/}}, \bibinfo{year}{2016}.
%Type = Article
\bibitem[{Wyss et~al.(2006)Wyss, Blair, Morris, Stone, and Weitz}]{Wyss2006a}
\bibinfo{author}{H.~M. Wyss}, \bibinfo{author}{D.~L. Blair},
  \bibinfo{author}{J.~F. Morris}, \bibinfo{author}{H.~A. Stone},
  \bibinfo{author}{D.~A. Weitz},
\newblock \bibinfo{title}{{Mechanism for clogging of microchannels}},
\newblock \bibinfo{journal}{Physical Review E - Statistical, Nonlinear, and
  Soft Matter Physics} \bibinfo{volume}{74} (\bibinfo{year}{2006})
  \bibinfo{pages}{061402}.
%Type = Misc
\bibitem[{iot(2016)}]{iota300}
\bibinfo{title}{{ECOM Iota 300 HPLC pump}},
  \bibinfo{howpublished}{{http://www.ecomsro.com/FS/0001-Ecom/files/products/Info-Iota50-100-300-pump-en.pdf}},
  \bibinfo{year}{2016}.
%Type = Misc
\bibitem[{hub(2016)}]{huber}
\bibinfo{title}{{Huber Petite Fleur w}},
  \bibinfo{howpublished}{{http://www.huber-online.com}}, \bibinfo{year}{2016}.
%Type = Misc
\bibitem[{cor(2016)}]{cori}
\bibinfo{title}{{Bronkhorst miniCori Flow}},
  \bibinfo{howpublished}{{http://www.bronkhorst.ch}}, \bibinfo{year}{2016}.
%Type = Misc
\bibitem[{swa(2016)}]{swage}
\bibinfo{title}{{Swagelok pressure transducer}},
  \bibinfo{howpublished}{{http://www.swagelok.com}}, \bibinfo{year}{2016}.
%Type = Misc
\bibitem[{val(2016)}]{valve}
\bibinfo{title}{{Swagelok valve SS-ORS2}},
  \bibinfo{howpublished}{{http://www.swagelok.com}}, \bibinfo{year}{2016}.
%Type = Misc
\bibitem[{fil(2016)}]{filter}
\bibinfo{title}{{Swagelok filter SS-2TF-15}},
  \bibinfo{howpublished}{{http://www.swagelok.com}}, \bibinfo{year}{2016}.
%Type = Misc
\bibitem[{nan(2016)}]{nanoports}
\bibinfo{title}{{NanoPort Assemblies}},
  \bibinfo{howpublished}{{https://www.idex-hs.com}}, \bibinfo{year}{2016}.
%Type = Misc
\bibitem[{rs(2016)}]{rs}
\bibinfo{title}{{rs-pro 4inch heater}},
  \bibinfo{howpublished}{{http://www.rs-online.com}}, \bibinfo{year}{2016}.
%Type = Misc
\bibitem[{var(2016)}]{variocam}
\bibinfo{title}{{infrared VarioCAM HD}},
  \bibinfo{howpublished}{{http://www.infratec.de}}, \bibinfo{year}{2016}.

\end{thebibliography}

%% Authors are advised to submit their bibtex database files. They are requested to list a bibtex style file in the manuscript if they do not want to use model1a-num-names.bst.



%% \bibitem{key}...
%%


% \end{thebibliography}

\end{document}